%% file: main.tex
\documentclass[final,5p,times,twocolumn]{elsarticle}

\usepackage{amssymb}
\usepackage{amsmath}

\usepackage[numbers]{natbib}
\usepackage{listings}
\usepackage{graphicx}

\usepackage{algorithm}
\usepackage{algorithmic}
\usepackage[shortlabels]{enumitem} 
\usepackage{xcolor}
\usepackage{subcaption}
\usepackage{float}
\usepackage{url}
\usepackage{hyperref}
\usepackage{titlesec}

\usepackage[acronym,]{glossaries} 

\input{glossaries}
\glsdisablehyper

\journal{Journal of Parallel and Distributed Computing}

\begin{document}
\begin{frontmatter}

\begin{graphicalabstract}
{
\centering
\includegraphics[width=1.93\linewidth]{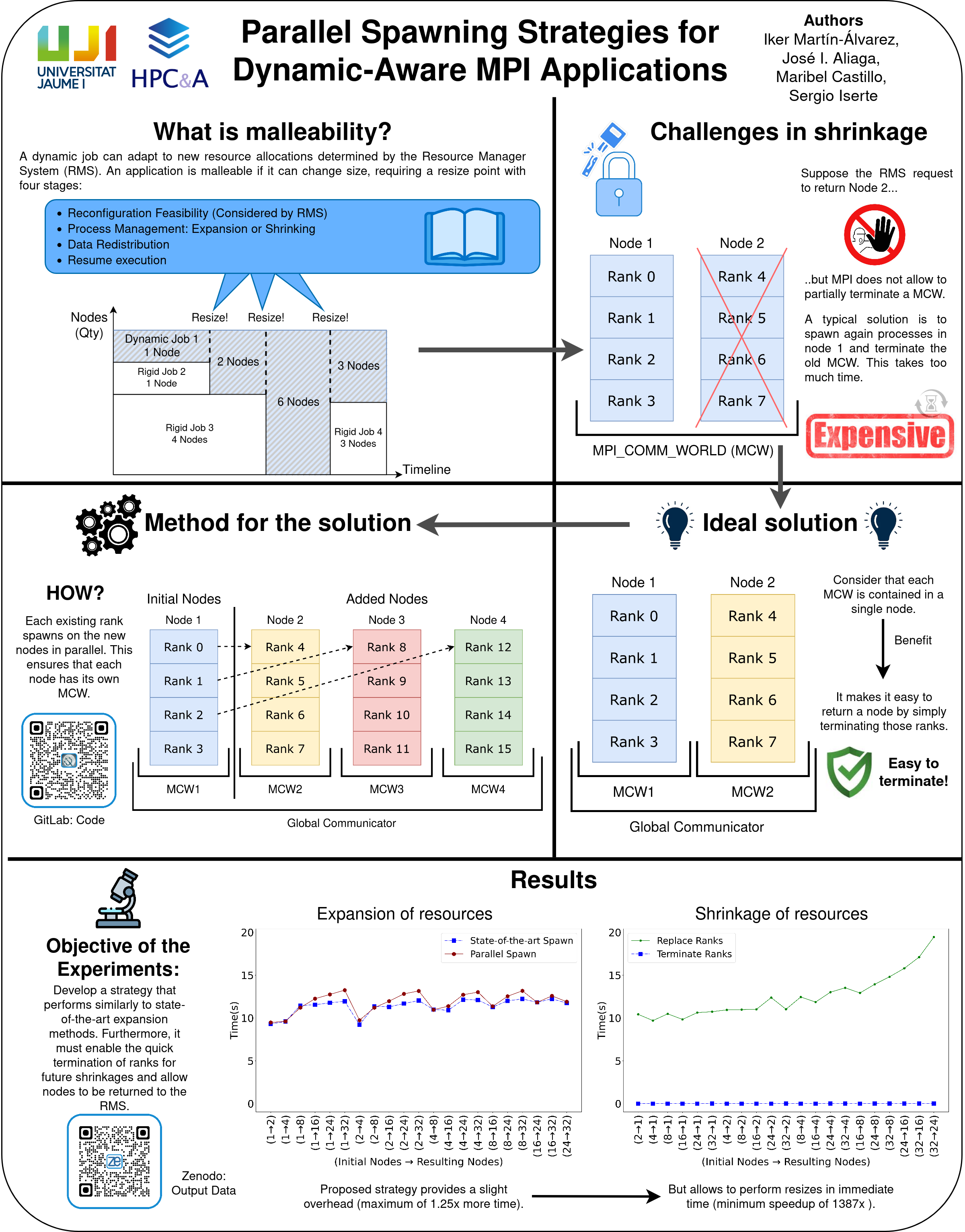}
}
\end{graphicalabstract}


\title{Parallel Spawning Strategies for Dynamic-Aware MPI Applications}

\author[1]{Iker Martín-Álvarez\corref{cor1}}
\ead{martini@uji.es}
\author[1]{José I. Aliaga}
\ead{aliaga@uji.es}

\author[1]{Maribel Castillo}
\ead{castillo@uji.es}


\affiliation[1]{organization={Dpto. de Ingeniería y Ciencia de los Computadores, Universitat Jaume I},
                addressline={Av. Vicent Sos Baynat, s/n }, 
                city={Castelló},
                postcode={12071}, 
                state={Comunitat Valenciana},
                country={Spain}}

\cortext[cor1]{Corresponding author}

\begin{abstract}
Dynamic resource management is an increasingly important capability of High Performance Computing systems, as it enables jobs to adjust their resource allocation at runtime. 
This capability can reduce workload makespan, substantially decreasing job waiting times and optimizing resource allocation.  
In this context, malleability refers to the ability of applications to adapt to new resource allocations during execution. 
Although beneficial, malleability incurs significant reconfiguration costs, making the reduction of these costs an important research topic.
Some existing solutions for MPI applications respawn the entire application, which is an expensive solution that avoids the reuse of original processes.
Other MPI solutions reuse them, but fail to fully release unneeded processes when shrinking, since some ranks within the same communicator remain active across nodes, preventing the application from returning those nodes to the system. 
This work overcomes both limitations by proposing a novel parallel spawning strategy, in which all processes cooperate in the spawning.
This allows expansions to reuse processes while also terminating unneeded ones.
This strategy has been validated on two systems with either machines with equal or different numbers of cores.
Experiments show that this strategy preserves competitive expansion times with at most a $1.13\times$ and $1.25\times$ overhead for equal and different number of cores per node, respectively. More importantly, it enables fast shrink operations that reduce their cost by at least $1387\times$ and $20\times$ in the same scenarios. 
\end{abstract}


\end{frontmatter}

\setlist[itemize]{noitemsep, topsep=2pt}
\setlist[enumerate]{noitemsep, topsep=2pt}



\input{glossaries_config}
\input{Sources/1-Intro-Mar}
\input{Sources/2-Background}
\input{Sources/3-SpawnBasic}
\input{Sources/4-SpawnMultiple}
\input{Sources/5-Results}
\input{Sources/6-Conclusions}

\titlespacing*{\section}{10pt}{6pt}{4pt}

\section*{Acknowledgements}
We would like thank Sergio Iserte for his valuable contributions to this work, his insightful discussions on possible improvements of the model, as well as his support in gaining access to the \gls{mn5} supercomputing facilities, which enabled the large-scale simulations presented in this article.

\section*{Declaration of competing interest}
The authors have no conflicts of interest to declare that are relevant to the content of this article.
\section*{Data availability}
The source codes for MaM are publicly available (\href{https://lorca.act.uji.es/gitlab/martini/malleability_benchmark/-/tree/Paper-ParallelSpawn}{https://lorca.act.uji.es/gitlab/martini/malleability\_benchmark/-/tree/Paper-ParallelSpawn}). All data used to create tables and figures in the Results section are available at~\cite{martin_alvarez_2025_17315810}.
\section*{Funding sources}
The researchers from UJI have been funded by the project PID2023-146569NB-C22, supported by MCIN\slash AEI\slash 10.13039\slash 501100011033.
Researcher I.~Martín-Álvarez was supported by the predoctoral fellowship ACIF\slash 2021\slash 260 from Valencian Region Government and European Social Funds.
%
\section*{CRediT authorship contribution statement}
\begin{itemize}
\item Iker Martín-Álvarez: Conceptualization, Methodology, Software, Investigation, Validation, Formal analysis, Data Curation, Visualization, Writing - Original Draft.
\item José I. Aliaga: Methodology, Writing - Review \& Editing, Supervision, Resources, Funding acquisition.
\item Maribel Castillo: Methodology, Formal analysis, Writing - Review \& Editing, Supervision.
\end{itemize}
\section*{Declaration of generative AI use}
During the preparation of this work the authors used ChatGPT exclusively to refine the language and improve textual clarity. After using this tool, the authors reviewed and edited the content as needed and take full responsibility for the content of the published article.
%
%
\bibliographystyle{elsarticle-num}
\bibliography{bib}
\end{document}

%% file: glossaries.tex
\newacronym{drm}{DRM}{Dynamic Resource Management}
\newacronym{hpc}{HPC}{High Performance Computing}
\newacronym{mp}{MP}{Message Passing}
\newacronym{mpi}{MPI}{Message Passing Interface}
\newacronym{rms}{RMS}{Resource Manager System}
\newacronym{slurm}{Slurm}{Simple Linux Utility for Resource Management}
\newacronym{io}{I/O}{Input/Output}
\newacronym{cr}{C/R}{Checkpoint and Restart}
\newacronym{ram}{RAM}{Random Access Memory}
\newacronym{ulfm}{ULFM}{User Level Failure Mitigation}
\newacronym{pmi}{PMI}{Process Management Interface}
\newacronym{pmix}{PMIx}{Process Management Interface - Exascale}
\newacronym{api}{API}{Application Programming Interface} 
\newacronym{spmd}{SPMD}{Single Program Multiple Data}
\newacronym{pmpi}{PMPI}{MPI Profiling Interface}
\newacronym{cg}{CG}{Conjugate Gradient}
\newacronym{pbicg}{PBiCGStab}{Preconditioned Biconjugate Gradient Stabilized}
\newacronym{mpdata}{MPDATA}{Multidimensional Positive Definite Advection Transport Algorithm}
\newacronym{pue}{PUE}{Power Usage Effectiveness}
\newacronym{dmr}{DMR}{Dynamic Management of Resources}
\newacronym{dmrlib}{DMRlib}{Dynamic Management of Resources library}
\newacronym{dmrapi}{DMR API}{Dynamic Management of Resources Application Interface}
\newacronym{gpu}{GPU}{Graphics Processing Unit}
\newacronym{cpu}{CPU}{Central Processing Unit}
\newacronym{ampi}{AMPI}{Adaptive MPI} 
\newacronym{pgas}{PGAS}{Partitioned Global Address Space}
\newacronym{apgas}{APGAS}{Asynchronous Partitioned Global Address Space}
\newacronym{smp}{SMP}{Symmetric MultiProcessor}
\newacronym{talp}{TALP}{Tracking Application Live Performance}
\newacronym{qos}{QoS}{Quality of Service}
\newacronym{mcw}{MCW}{\texttt{MPI\_COMM\_WORLD}}
\newacronym{dpp}{DPP}{Dynamic Processes with Process Sets}
\newacronym{pset}{PSet}{Process Sets}
\newacronym{col}{COL}{Cooperative Optimization Language}
\newacronym{prrte}{PRRTE}{PMIx Reference RunTime Environment}
\newacronym{posix}{POSIX}{Portable Operating System Interface}
\newacronym{pid}{PID}{Process IDentifier}
\newacronym{ofi}{OFI}{Open Fabric Interface}
\newacronym{ucx}{UCX}{Unified Communication X}
\newacronym{edr}{EDR}{Endpoint Detection and Response}
\newacronym{ndr}{NDR}{Network Detection and Response}
\newacronym{rma}{RMA}{Remote Memory Access}
\newacronym{mam}{MaM}{Malleability Module}
\newacronym{sam}{SAM}{Synthetic Application Module}
\newacronym{map}{MaP}{Malleability Point}
\newacronym{mn5}{MN5}{Marenostrum 5}
\newacronym{libadr}{libADR}{Library of Awesome Dynamic Resources}

\newacronym{nc}{NC}{Number of Children processes}
\newacronym{np}{NP}{Number of Parent processes}
\newacronym{ns}{NS}{Source processes}
\newacronym{nt}{NT}{Target processes}
\newacronym{nd}{ND}{Number of Drains processes}
\newacronym{kb}{KB}{KiloBytes}
\newacronym{mb}{MB}{MegaBytes}
\newacronym{gb}{GB}{GigaBytes}
\newacronym{gbi}{Gbit}{Gigabits}
\newacronym{h0}{H0}{Null Hypothesis} 
\newacronym{avg}{AVG}{Average}
\newacronym{std}{STD}{STandard Deviation}

\newacronym{zs}{ZS}{Zombies Shrinkage}
\newacronym{ts}{TS}{Termination Shrinkage}
\newacronym{ss}{SS}{Spawn Shrinkage}

\newacronym{srs}{SRS}{Stop Restart Software}
\newacronym{rss}{RSS}{Runtime Support System}
\newacronym{ibp}{IBP}{Internet Backplane Protocol}
\newacronym{pcm}{PCM}{Process Checkpointing and Migration}
\newacronym{ios}{IOS}{Internet Operating System}
\newacronym{scr}{SCR}{Scalable Checkpoint and Restart}
\newacronym{ers}{ERS}{Elastic Runtime Scheduler}
\newacronym{abs}{ABS}{Adaptive Batch Scheduler}
\newacronym{mtct}{MTCT}{MPI to Compute Time}
\newacronym{rapl}{RAPL}{Running Average Power Limit}
\newacronym{blas}{BLAS}{Basic Linear Algebra Subprograms}
\newacronym{blacs}{BLACS}{Basic Linear Algebra Communication Subprograms}
\newacronym{blis}{BLIS}{BLAS-like Library Instantiation Software}
\newacronym{cpm}{CPM}{Computational Prediction Model}
\newacronym{ccs}{CCS}{Converse Client-Server}
\newacronym{tcp}{TCP}{Transmission Control Protocol}
\newacronym{ip}{IP}{Internet Protocol}
\newacronym{ams}{AMS}{Application Management System}
\newacronym{lam}{LAM}{Local Area Multicomputer}
\newacronym{cfd}{CFD}{Computational Fluid Dynamics}
\newacronym{ic}{IC}{Intelligent Controller}
\newacronym{hdfs}{HDFS}{Hadoop Distributed File System}
\newacronym{mape}{MAPE}{Monitor–Analyze–Plan–Execute}
\newacronym{limitless}{LIMITLESS}{Light-weight MonItoring Tool for LargE Scale Systems}
\newacronym{pop}{POP}{Performance Optimisation and Productivity}
\newacronym{drom}{DROM}{Dynamic Resource Ownership Management}
\newacronym{dlb}{DLB}{Dynamic Load Balancing}
\newacronym{lewi}{LeWI}{Lend When Idle}
\newacronym{sdpolicy}{SD-Policy}{Slowdown-Driven Policy}
\newacronym{parsl}{Parsl}{Parallel Scripting Library}
\newacronym{dfk}{DFK}{DataFlowKernel}
\newacronym{htex}{HTEX}{High Throughput Executor}
\newacronym{dvm}{DVM}{Distributed Virtual Machine}
\newacronym{json}{JSON}{JavaScript Object Notation}
\newacronym{fifo}{FIFO}{First-In, First-Out}
\newacronym{fcfs}{FCFS}{First-Come, First-Served}
\newacronym{cmb}{CMB}{Communication Message Broker}
\newacronym{kvs}{KVS}{Key-Value Store}
\newacronym{unix}{UNIX}{UNiplexed Information Computing System}
\newacronym{compss}{COMPSs}{COMP Superscalar}
\newacronym{bsc}{BSC}{Barcelona Supercomputing Center}
\newacronym{pdr}{PDR}{Parallel Distributed Runtime}
\newacronym{parm}{PARM}{Power-Aware Resource Manager}
\newacronym{openmp}{OpenMP}{Open Multi-Processing}
\newacronym{ompt}{OMPT}{OpenMP Toolkit}
\newacronym{ear}{EAR}{Energy Aware Runtime}

\newacronym{Edrm}{DRM}{Gestión de Recursos Dinámicos}
\newacronym{Ehpc}{HPC}{Computación de Altas Prestaciones}
\newacronym{Erms}{RMS}{Sistema de Gestión de Recursos}
\newacronym{Empi}{MPI}{Interfaz de Paso de Mensajes}

%% file: glossaries_config.tex
\glsunset{api}
\glsunset{unix}
\glsunset{cpu}
\glsunset{gpu}
\glsunset{ofi}
\glsunset{ucx}
\glsunset{ndr}
\glsunset{pid}
\glsunset{kb}
\glsunset{mb}
\glsunset{gb}
\glsunset{gbi}
\glsunset{avg}
\glsunset{std}
\glsunset{nc}
\glsunset{np}
\glsunset{slurm}
\glsunset{mpi}
\glsunset{io}
\glsunset{dmr}
\glsunset{dpp}
\glsunset{libadr} 
\glsunset{mam}
\glsunset{dmrapi}

%% file: Sources/1-Intro-Mar.tex
\section{Introduction}
\label{sec:5-intro}
Despite advances in \gls{hpc} systems during the exascale era, the efficient utilisation of resources such as \gls{cpu}s, \gls{gpu}s and memory remains a significant challenge~\cite{Jie2023}. Scientific applications often underuse their allocations due to factors such as overcommitment, bottlenecks or limited scalability. 

To address this issue, \gls{rms} must be able to distribute resources dynamically, a capability known as \gls{drm}. This capability enables \gls{rms}s to adjust job allocations at runtime, provided that the applications are malleable, that is, capable of resizing and adapting to changes in resource availability during execution. Such flexibility allows applications to achieve more effective optimization of resource utilization~\cite{Chadha2021}, throughput~\cite{sergiothesis}, energy efficiency~\cite{Cascajo2023,Iserte2019a}, and \gls{io} performance~\cite{Sanchez2023}.

A prior study by the US Exascale Computing Project~\cite{project-exascale} suggest that \gls{drm} could significantly enhance the performance of both applications and systems, making it a key contribution for improving relevant metrics in future platforms~\cite{osti_1222713}. 
However, its adoption remains limited compared to other \gls{mpi} features~\cite{HORI2021102853}, primarily due to the complexity involved in adapting applications to support malleability and enabling systems to dynamically reallocate resources at runtime. 

There are already libraries available to perform reconfigurations in applications; some of the most relevant are \gls{dpp}\cite{Dominik22}, \gls{dmr}\cite{sergiothesis}, FlexMPI\cite{Martin2013}, or \gls{mam}\cite{mam_api_2024}. A particularly challenging scenario for most of these tools is shrinking, i.e., reducing the number of active processes in a job at runtime to free resources or adapt to changing workload demands. This operation is limited by the rigid structure of the standard \gls{mpi} communicator \gls{mcw}, which does not support the termination of a subset of ranks, even if separated with a \texttt{MPI\_Comm\_split} function.

In the paper~\cite{spawn_methods} two methods, Baseline and Merge, are evaluated with different strategies. The latter method allowed to shrink the application by the conversion of processes into zombies, which we denote as \gls{zs}. The \gls{zs} offered lower reconfiguration times than traditional \gls{ss} obtained with the Baseline method. Nevertheless, a major limitation of \gls{zs} is the persistence of zombie processes, ranks that remain asleep until the application terminates, as they cannot be fully removed from the global \gls{mcw}. This creates an issue: the nodes on which these zombies are mapped cannot be returned to the \gls{rms} until they terminate, preventing node reuse. 
A common workaround involves isolating each \gls{mcw} communicator on a separate node, allowing finer control over process groups. However, this solution depends on node-specific spawning, which suffers from poor scalability due to its inherently sequential nature \cite{Flex_mpi_spawning}. Yet, by using this technique it is possible to shrink the application by terminating those processes, an approach that we denote as \gls{ts}, which overcomes the key limitation of \gls{zs}, its inability to fully terminate processes and make them persist as zombies.

To overcome the scalability constraints identified in~\cite{Flex_mpi_spawning}, we propose a novel technique that preserves scalability during process spawning while enabling the use of \gls{ts} for shrinkages. Our technique also isolates each \gls{mcw} communicator on a separate node; however, instead of performing this step sequentially, it is carried out in parallel, allowing each available process to initiate an independent spawn. Additionally, newly spawned processes participate in subsequent spawn operations within the same reconfiguration if further spawning is required.

These improvements make our solution particularly suitable for \gls{drm} in large-scale \gls{hpc} environments, with full support for nodes with different numbers of cores, and for jobs with different numbers of cores on each node. Consequently, to evaluate the novel technique, this work considers two types of node allocations: homogeneous, where jobs use the same number of cores in each node, and heterogeneous, where the number of cores used in each node can differ. Additionally, the improvements can be used for oversubscription in both types of allocation, where the number of processes exceeds the number of cores allocated to the job. Nevertheless, it is not considered the challenge of load balancing between different types of hardware, which is postponed as future work. 

This approach has been included in the \gls{mam} library~\cite{mam_api_2024}, which integrates multiple techniques for resizing and data redistribution, allowing the selection of the optimal solution depending on the context. The main contributions of this work are:
\begin{itemize}
\item The design and implementation of a novel parallel spawning algorithm tailored to the \gls{mpi} model, enabling scalable malleable executions for \gls{hpc} applications in homogeneous node allocations. 
\item An adaptation of the previous technique to enable resizing in heterogeneous node allocations.
\item The capability to shrink malleable applications by terminating processes (\gls{ts}) without spawning new ones (\gls{ss}) or leaving them as zombies (\gls{zs}).
\item A comprehensive performance evaluation of the proposed algorithms on jobs with homogeneous node allocations, demonstrating their advantages.
\item A performance evaluation of the extended algorithm for jobs with heterogeneous node allocations, demonstrating its advantages.
\end{itemize}

The remainder of this paper is organized as follows: Section~\ref{sec:5-background} presents a comprehensive review of related work on \gls{mpi} malleability and spawning techniques. 
Section~\ref{sec:5-spawn_basics} provides an overview of process spawning mechanisms in \gls{mpi} and their integration into \gls{mam}.
Section~\ref{sec:5-method} explains the two new parallel spawning algorithms proposed in this work. 
Section~\ref{sec:5-results} presents an evaluation of their performance. Finally, Section~\ref{sec:5-conclusions} concludes the paper and outlines directions for future work.

%% file: Sources/2-Background.tex
\section{Related Work}
\label{sec:5-background}
According to the classification proposed by Feitelson and Rudolph~\cite{Feitelson1996}, applications executed in large-scale computing facilities can be grouped into four categories (see Table~\ref{tab:5-job-classification}).
This classification is based on two main criteria: (1) whether the number of processes can change during execution, i.e., whether reconfiguration is supported, and (2) who determines the new job size, either the application or the RMS.

\begin{table*}[tb!]
\centering
\begin{tabular}{|l|c|c|p{7.5cm}|}
\hline
\textbf{Job Type} & \textbf{Resource Allocation} & \textbf{Who Sets Size} & \textbf{Description} \\
\hline
\textbf{Rigid Job} & Static & Application & Can only be executed with a fixed number of processes defined by the user. No reconfiguration is allowed. \\
\hline
\textbf{Moldable Job} & Static & \gls{rms} & Can start with a variable number of processes. The \gls{rms} determines the size when the execution begins. \\
\hline
\textbf{Evolving Job} & Dynamic & Application & Includes a user-defined reconfiguration scheme. The job will only continue if the \gls{rms} satisfies the resource request; otherwise, it waits or it is cancelled. \\
\hline
\textbf{Malleable Job} & Dynamic & \gls{rms} & Can be reconfigured at runtime if the \gls{rms} decides to adjust the resource allocation. \\
\hline
\end{tabular}
\caption[Classification of parallel jobs according to Feitelson and Rudolph]{Classification of parallel jobs according to Feitelson and Rudolph~\cite{Feitelson1996}.}
\label{tab:5-job-classification}
\end{table*}

In this paper, we consider malleability as the ability of a distributed parallel job to change its size in terms of \gls{mpi} ranks by adjusting the computational resources initially allocated to the job at any point during execution as often as required. 
This capability is activated at specific checkpoints within the application and requires the execution of a series of stages:
\begin{enumerate}
    \item \textbf{Reconfiguration feasibility}: 
    The \gls{rms} determines whether the job should be resized according to a dynamic resource allocation policy. If not, the subsequent steps are not performed.
    
    \item \textbf{Process management}. 
    The \gls{rms} assigns a new number of cores to the job, which determines how many \gls{mpi} ranks should be created or terminated. 
    Processes before resizing are considered \textit{sources}, while those that continue after resizing are considered \textit{targets}.
    
    \item \textbf{Data redistribution}: 
    \textit{Sources} transfer their data to \textit{targets}.
    
    \item \textbf{Resume execution}. The application resumes execution with the \textit{targets}.
\end{enumerate}

Incorporating all these stages into a parallel application is a complex task that goes far beyond simple integration with the \gls{rms}. It requires modifications to the application's structure to support dynamic creation and termination of processes, the efficient management of data redistribution, and the reconfiguration of communication patterns on the fly. These challenges make dynamic applications difficult to be implemented from scratch, since they often require dedicated tools and frameworks to simplify the transformation of traditional applications into dynamic ones. Without such support, adapting applications to handle resizing becomes both error-prone and time-consuming.

One malleable solution is Flex-MPI~\cite{Martin2013}, a library built on top of MPICH~\cite{mpich_website} that bases its spawn operations on the \texttt{MPI\_Comm\_spawn} function. 
The primary goal of this solution is to help applications become dynamic and the main method to resize in Flex-MPI is the Merge method, which is described below.
When an application expands from $x$ processes to $y$ processes, only the additional $y-x$ processes are spawned and added to the existing ones. 
Then, each new process is created through a separate call to the spawn function, which allows a fine grain when shrinking the application, as dynamic processes can be terminated at any time. 
Nevertheless, there is a limitation with the minimum number of ranks: the initial number of ranks cannot be terminated until the end of the application.

A similar solution is \gls{mam}~\cite{mam_api_2024}, a malleability module also built on top of MPICH and within the \gls{mpi} scope, developed to analyze and compare different malleability techniques. 
Like Flex-MPI, it bases its spawn operation on the \texttt{MPI\_Comm\_spawn} function and
implements Baseline and Merge methods, in which the former
always spawns the new number of processes $y$ and terminates
the previous $x$ processes.
\gls{mam} is part of the Proteo framework~\cite{proteo_2024}, which helps transform applications into malleable ones. It supports the generation of dynamic workloads, and enables performance analysis of different resize operations to select the optimal alternative for a given situation, by simulating different scenarios.

Another approach is \gls{dpp}~\cite{Dominik22}, which explores the new concept of \gls{mpi} Sessions, extending it to implement on-the-fly dynamic mechanisms. 
Since the sessions model operates outside the traditional \gls{mpi} environment, the \gls{mcw} communicator does not exist and some functions, such as \texttt{MPI\_Comm\_spawn}, cannot be used. 
However, alternative non-standard \gls{mpi} functions are provided to allow communication between ranks. 
\gls{dpp} uses a custom dynamic \gls{rms}, which determines whether a job should get or return resources.
When a job gets additional resources, it notifies a \gls{prrte} daemon, which then spawns the new processes.
Once the processes are ready, the job is notified of the new resources and how to connect with them, which is managed by \gls{dpp}. 
During shrink, processes do not have the limitations of the traditional environment and can be terminated arbitrarily.

Lastly, an alternative solution is \gls{dmrlib}~\cite{sergiothesis}, which consists of two components: a parallel distributed runtime based on \gls{mpi} and an extension of \gls{slurm} to enable the execution of malleable jobs. 
Initially, the library only allowed the simplest approach, the Baseline method.
However, the library now supports other environments for process management, such as \gls{mam}~\cite{ISERTE2025a} or \gls{dpp}~\cite{Dominik25}, providing a wider range of possibilities.


For further details on \gls{mpi} spawn based malleability, see~\cite{State22}, while a broader review of dynamic resource management can be found in~\cite{State24}.

%% file: Sources/3-SpawnBasic.tex
\section{MaM overview}
\label{sec:5-spawn_basics}

\gls{mam} is the Proteo module responsible for completing malleability. 
It provides a set of features designed to transform traditional \gls{mpi} applications into malleable ones, requiring minimal effort from developers.
Additionally, it defines a simple and intuitive interface, which makes easier to integrate malleability into existing parallel applications~\cite{mam_api_2024}.

Specifically, \gls{mam} implements multiple methods and strategies to handle the most critical stages of malleability, process management and data redistribution. These techniques allow applications to dynamically adjust the number of \gls{mpi} ranks during execution. Using \gls{mam} to incorporate malleability into an application requires specifying a method with none, one, or more of the strategies for process management and data redistribution. 

For the process management stage, i.e., resizing from an initial group of \textit{\gls{ns}} to a new group of \textit{\gls{nt}}, \gls{mam} offers two main methods: the Baseline method, which always creates a new group of processes, and the Merge method, which reuses existing processes and only spawns or terminates the difference in ranks. 
These methods can be combined with two strategies: the Asynchronous strategy, which overlaps process creation with application execution to reduce downtime, and the Single strategy, where only one process performs the spawn operation and informs the rest afterwards. 
This flexible design allows users to decide the preferred configuration according to their application needs.



%% file: Sources/4-SpawnMultiple.tex
\section{Strategies for spawning processes in parallel}
\label{sec:5-method}
As previously discussed, the \gls{zs} approach is constrained by its inability to fully terminate unnecessary processes during execution. As a consequence, it is not possible to return the corresponding nodes until \gls{mcw}, to which they belong, has terminated completely. However, a previous study~\cite{spawn_methods} using a test system showed for shrinkages that the \gls{zs} method delivered a speedup of $36\times$ over \gls{ss} methods. Consequently, reaching a solution for shrinkage with TS methods would prove a valuable advantage to reduce the overhead caused by resizing. To overcome these challenges, this work proposes new strategies based on parallel spawning, which allows processes to be resized efficiently and create an \gls{mcw} for each node, facilitating the use of a \gls{ts} approach.

Parallel process creation is inherently complex and is coordinated across multiple phases: (1) New processes groups are spawned and communication ports are opened. (2) This is followed by a synchronization phase, where all groups wait until every required port is ready. (3) Next, the newly spawned groups are merged into a single communicator through the ports. (4) Finally, a global reordering of ranks is performed to maintain a logical order between nodes.

The first phase has been addressed through two alternative approaches: A simpler option only for homogeneous node allocations called Hypercube strategy, and an improved but more complex one that also works for jobs with heterogeneous node allocations, called Iterative Diffusive strategy. The following subsections detail these approaches, and explain how the four phases are optimized, highlighting the specific operations involved in each step.

\subsection{Hypercube Strategy}
\label{subsec:5-hypercube}

This parallel strategy is based on invoking the \texttt{MPI\_Comm\_spawn} function once for each node to be used (up to the total number of required \gls{mpi} ranks), assigning all available cores of the node (process group) to each newly spawned process.

The procedure is organized into a series of steps. In each step, all existing processes concurrently execute a spawn operation, each one acting on a different node. As the number of spawned processes increases, so does the number of processes participating in the spawning operation. Consequently, with each successive step, a growing number of processes are involved in initiating new spawns, enabling a scalable and distributed expansion across nodes.

In the first step, only the \textit{source} processes participate in the operation. In total, up to $C \cdot I$ processes can be launched, where $C$ is the number of cores per node and $I$ is the number of initial nodes. From the second step onward, both the original and previously spawned processes participate, exponentially increasing the spawning capacity. However, in the final step, only the first $x$ processes participate, where $x$ corresponds to the remaining number of nodes needed to reach the desired number of nodes.

As an example, consider a system with $20$ cores per node. Starting with a single fully occupied node, the first step could launch up to $20$ additional nodes. In the second step, with $420$ processes available ($21$ nodes with $20$ cores each), it would be possible to occupy $420$ additional nodes, creating $20$ processes per node.

From a technical standpoint, this strategy uses \texttt{MPI\_Comm\_spawn} in combination with the \texttt{MPI\_COMM\_SELF} communicator and a \texttt{MPI\_Info} object that specifies the target node for each new group of processes. Furthermore, all processes in each node are assigned a group identifier ranging from $0$ to the total number of groups to be spawned (\textit{group\_id}), identifying which group it belongs.

From a mathematical perspective, the strategy can be described as a geometric growth, where each existing process expands into $C$ new nodes while the process that creates each node remains active. Therefore, the effective growth factor per step is $C+1$. The initial number of nodes, $I$, is determined by the total number of \textit{source} processes \gls{ns} as:
$$
I = \frac{NS}{C}\quad,\quad NS\;mod\;C = 0.
$$ 
Using this initial condition and the growth factor, the total number of nodes at step $s$ is given by:
%
\begin{equation}
    T_{s} = 
    \begin{cases} 
    (C+1)^s \cdot I - I, & \text{if Method} = \text{Baseline} \\
    (C+1)^s \cdot I, & \text{if Method} = \text{Merge}
    \end{cases} \quad ,
    \label{equation:5-max_nodes}
\end{equation}
where $T_s$ represents the total nodes at step $s$ and $C+1$ is the growth factor including the cores and the process that spawns them. Additionally, it considers the chosen spawning method, either Baseline or Merge. For the former, the initial nodes ($I$) are not considered to reach the target nodes ($N$), as the source processes will be removed after the reconfiguration. Thus, the corresponding total number of processes at step $s$ is simply obtained by multiplying the number of nodes by the cores per node:
\begin{equation}
{t_s} = C \cdot T_s.
\label{equation:5-max_procs}
\end{equation}
Finally, the total number of steps required to reach a target of $N$ nodes can be derived from Equation~\ref{equation:5-max_nodes} by solving for $s$ as:
\begin{equation}
s = \left\lceil\frac{\ln(N/I)}{\ln(C+1)}\right\rceil,
\label{equation:5-max_steps}
\end{equation}
where, $N$ represents the desired total number of nodes, calculated as:
$$
N = \frac{NT}{C}\quad,\quad NT\;mod\;C = 0,
$$ 
%
where, \gls{nt} is the total number of \textit{target} processes.
 
\begin{figure}[tb!]
\centering
\includegraphics[width=0.6\linewidth]{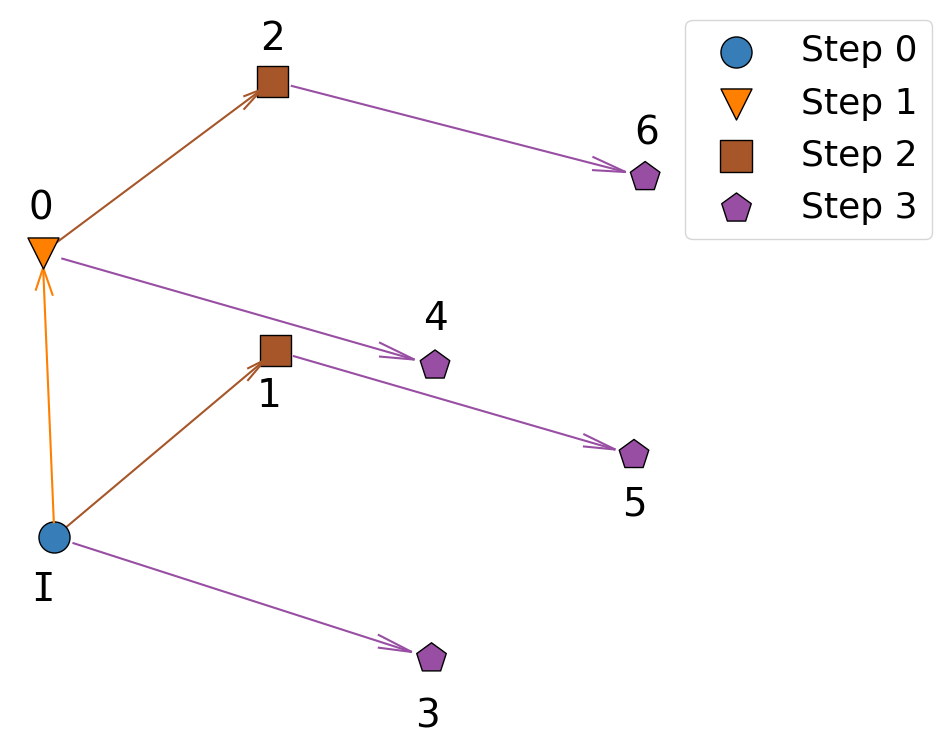}
\caption[Generation of $7$ groups using parallel spawning]{Generation of $7$ groups using parallel spawning. Each step is represented with different markers, and the number marks the group identifier.}
\label{fig:5-Parallel_Spawn-8Nodes}
\end{figure}

Figure~\ref{fig:5-Parallel_Spawn-8Nodes} illustrates this strategy with a simple example in which the number of processes is expanded from $NS = 1$ to $NT = 8$, assuming an architecture with $C = 1$ core per node.
In this case, seven additional groups are spawned over the course of three steps. The process is visualized as a cube, where the edges indicate which group spawns which in each step, and the vertices are labeled with the corresponding \texttt{group\_id}, or with the label $I$ for the initial group. By the end of the procedure, a total of eight nodes are occupied.

This strategy removes the shared \gls{mcw} when expanding across multiple nodes. Therefore, it enables shrinkage via the \gls{ts} method by releasing the excess processes.

\subsection{Iterative Diffusive Strategy}
\label{subsec:5-iterative_diff}
The main limitation of the Hypercube strategy is that all spawned process groups have a constant size. The Iterative Diffusive parallel strategy can use variable sizes for each process group. This makes it particularly well-suited for jobs with heterogeneous node allocations.

As in the previous scheme, the algorithm proceeds in discrete steps, where all existing processes collaborate to spawn new groups on additional nodes. The assignment of a unique group identifier ($group\_id$), ranging from $0$ to the total number of groups to be spawned, is also preserved.

Three vectors are needed to describe the allocation configuration, and are maintained through the execution. The number of entries in these vectors is equal to the number of nodes in the allocation ($N$).
\begin{itemize}
    \item $A$: Contains the total number of cores assigned to the job on each node.
    \item $R$: Stores the number of processes of the job running on each node.
    \item $S$: Keeps the number of processes of the job to be spawned in each node when executing this algorithm. This vector is computed from the previous ones as follows:
\end{itemize}
$$
S_i = A_{i} - R_i , \quad \text{for } i = 0, 1, \dots, N-1.
$$
Based on these vectors, a stepwise sequence can be constructed to determine the total number of processes and nodes involved at each spawning step.
Note that only nodes with positive entries in the vector $S$ will allocate new processes; thus, each entry greater than zero indicates the creation of a new group on the corresponding node. 
The number of existing processes in each step $s$ can then be calculated as follows:
\begin{equation}
    t_{s} = 
    \begin{cases} 
    \sum\limits_{j=0}^{N-1} R_{j}, & \text{if } s = 0 \\
    t_{s-1} + g_{s}, & \text{if } s > 0
    \end{cases} \quad ,
    \label{equation:5-hetero_total_procs}
\end{equation}
where $t_s$ is the total processes existing at the end of step $s$, calculated as a function of the cumulative total of previous steps ($t_{s-1}$) and the number of processes generated in the current step ($g_s$). The initial state ($s=0$) corresponds to the number of \textit{source} processes specified in vector $R$, and the number of generated processes ($g_s$) in a given step can be calculated as the sum of a sequence of elements in vector $S$:
\begin{equation}
    g_{s} = \sum\limits_{i=\lambda_{s-1}}^{\min\left(N,\, \lambda_{s}\right) - 1} S_{i}, 
    \label{equation:5-hetero_generated_procs}
\end{equation}
where the limits $\lambda_s$ defines the range of indices of vector $S$ used at each spawning step, as follows:
\begin{equation}
    \lambda_{s} = 
    \begin{cases} 
    0, & \text{if } s = 0 \\
    \lambda_{s-1} + t_{s-1}, & \text{if } s > 0
    \end{cases} \quad .
    \label{equation:5-hetero_indexes}
\end{equation}
%
These limits ensure that each step consumes a contiguous segment of $S$, without overlap or omission, while maintaining a sequential progression through the vector.

Specifically, $\lambda_{s-1}$ marks the starting index for step $s$, which corresponds to the total number of processes handled in all previous steps. Conversely, $\lambda_s - 1$ marks the last index to be included in the current step, extending over the $t_{s-1}$ processes that participate at this stage. As a result, each step $s$ uses exactly $t_{s-1}$ entries of $S$, starting right after the portion used in step $s - 1$. The minimum in the upper index for Equation~\ref{equation:5-hetero_generated_procs} is used to ensure no more than $N$ spawns are performed. In all equations, null elements in vector $S$ are disregarded.

The cumulative number of occupied nodes ($T_s$) up to step $s$ is calculated from these previous equations as follows:
\begin{equation}
    T_{s} = 
    \begin{cases}
    I, & \text{if } s = 0 \\
    T_{s-1} + G_{s}, & \text{if } s > 0
    \end{cases} \quad ,
    \label{equation:5-hetero_total_nodes}
\end{equation}
where for each step, the sum of the nodes from the previous step ($T_{s-1}$) and the nodes added during the current step ($G_{s}$) are calculated. 
In the first step ($s=0$) the initial number of nodes is used ($I$). For the next steps the number of nodes mainly depends on $G_{s}$, which represents the number of new nodes to add. A new node is added if the corresponding positions in the vectors $R$ and $S$ do not contain any processes before spawning ($R_i=0$) and if there are processes to spawn ($S_i > 0$). Therefore, one node is added for each position considered in the range defined by Equation~\ref{equation:5-hetero_indexes}, only if both conditions hold. This can be represented as follows:
\begin{equation}
    G_{s} = \sum\limits_{i=\lambda_{s-1}}^{\min\left(N,\, \lambda_{s}\right) - 1} \{R_i = 0 \wedge S_i > 0 \},
    \label{equation:5-hetero_generated_nodes}
\end{equation}
where the limits $\lambda_s$ are calculated by using Equation~\ref{equation:5-hetero_indexes}. When the condition of the summation is met, it adds 1, or 0 otherwise.

Table~\ref{tab:5-hetero_example} shows how the different Equations behave depending on the initial parameters $A$, $R$, $S$, $N$ and $I$. In this example, $s=0$ is the initial state before executing this strategy. 
\begin{table}[tb!]
\centering
\begin{tabular}{|crrrrr|}
\hline
\multicolumn{6}{|c|}{\textbf{Initial Vectors}} \\ \hline
\multicolumn{1}{|c|}{A} & \multicolumn{5}{c|}{{[}4, 2, 8, 12, 3, 3, 4, 4, 6, 3{]}} \\ \hline
\multicolumn{1}{|c|}{R} & \multicolumn{5}{c|}{{[}2, 0, 0, \hfill\:0, \:0, 0, 0, 0, 0, 0{]}} \\ \hline
\multicolumn{1}{|c|}{S} & \multicolumn{5}{c|}{{[}2, 2, 8, 12, 3, 3, 4, 4, 6, 3{]}} \\ \hline
\multicolumn{6}{|c|}{\textbf{Computed Values per Step}} \\ \hline
\multicolumn{1}{|c|}{$s$} & \multicolumn{1}{c|}{$t_s$} & \multicolumn{1}{c|}{$g_s$} & \multicolumn{1}{c|}{$\lambda_s$} & \multicolumn{1}{c|}{$T_s$} & \multicolumn{1}{c|}{$G_s$} \\ \hline
\multicolumn{1}{|r|}{0} & \multicolumn{1}{r|}{2} & \multicolumn{1}{r|}{-} & \multicolumn{1}{r|}{0} & \multicolumn{1}{r|}{1} & - \\ \hline
\multicolumn{1}{|r|}{1} & \multicolumn{1}{r|}{6} & \multicolumn{1}{r|}{4} & \multicolumn{1}{r|}{2} & \multicolumn{1}{r|}{2} & 1 \\ \hline
\multicolumn{1}{|r|}{2} & \multicolumn{1}{r|}{40} & \multicolumn{1}{r|}{34} & \multicolumn{1}{r|}{7} & \multicolumn{1}{r|}{8} & 6 \\ \hline
\multicolumn{1}{|r|}{3} & \multicolumn{1}{r|}{49} & \multicolumn{1}{r|}{9} & \multicolumn{1}{r|}{47} & \multicolumn{1}{r|}{10} & 2 \\ \hline
\end{tabular}
\caption[Iterative diffusive procedure example for a new allocation of resources]{Iterative diffusive procedure example for a new allocation of resources from $1$ node ($I$) to $10$ nodes ($N$).}
\label{tab:5-hetero_example}
\end{table}

Similar to the previous approach, this one ensures that no shared \gls{mcw} exists across the expanded nodes, with the added advantage of accounting for a heterogeneous number of cores on each expanded node. Moreover, it also enables fast shrinkage through \gls{ts} method.

\subsection{Synchronization Between Process Groups}
\label{subsec:5-synch_step}
The purpose of synchronizing the groups is to ensure that all processes know that every port is active before any connection attempt is made. Once all processes have been spawned, establishing these connections is essential, as they enable direct communication among them. Due to the chosen process creation strategies, each newly spawned group only knows its parent process (its creator) and, if applicable, its own child processes. Consequently, as shown in Figure~\ref{fig:5-Parallel_Spawn-8Nodes}, processes can only communicate directly with those linked to them by an edge in the spawn graph.

To enable communication between separate groups, the functions \texttt{MPI\_Comm\_connect} and \texttt{MPI\_Comm\_accept} are used. These functions establish an inter-communicator between two independent groups. However, to use them correctly, certain preliminary steps must be taken.

First, the process groups must be prepared for connection. Thus, the processes designated as roots in the \texttt{MPI\_Comm\_accept} calls must first open a port using \texttt{MPI\_Open\_port} and publish the service name via \texttt{MPI\_Publish\_name}. Since \texttt{MPI\_Comm\_connect} must be called after the corresponding port has been opened by the other group, a global synchronization phase between all groups is required. Otherwise, execution errors may occur, as observed with implementations such as MPICH\footnote{This behavior has only been tested with MPICH.}. For the rest of the paper, the root process of each group is always the process with rank $0$ in its communicator, which is required by the functions discussed here.

Each group carries out the synchronization using a dedicated subcommunicator consisting of three stages.
\begin{enumerate}[1]
\item \textbf{Subcommunicator creation}.
Each group creates a subcommunicator using \texttt{MPI\_Comm\_split}, including the root process of the group and all processes of the group that have spawned child groups. This subcommunicator will be used to coordinate intra-group synchronization.

\item \textbf{Upside Synchronization}.
Each process that has spawned child groups waits to receive a token from each of them using \texttt{MPI\_Irecv} and \texttt{MPI\_Waitall}. 
Then, all processes in the subcommunicator execute a call to \texttt{MPI\_Barrier}, ensuring that all tokens have been received.
Finally, the root process (if not part of the source group) notifies its parent group using \texttt{MPI\_Send}.

\item \textbf{Downside Synchronization}.
The root process of each group (except the source group) waits to receive a token from its creator via \texttt{MPI\_Recv}.
Then, processes in the subcommunicator perform a \texttt{MPI\_Barrier} call (excluding the source group).
Finally, each process that has spawned child groups sends them a token using \texttt{MPI\_Isend}, followed by \texttt{MPI\_Waitall}.
\end{enumerate}
The barriers included in the stages are essential to ensure that no group proceeds before all its descendants have completed the Upside step, and to prevent any group from notifying its children before being notified by its parent group. However, the \textit{Waitall} is required only to ensure the communication is completed.

\begin{figure*}[tb!]
  \centering
  \includegraphics[width=0.80\textwidth]{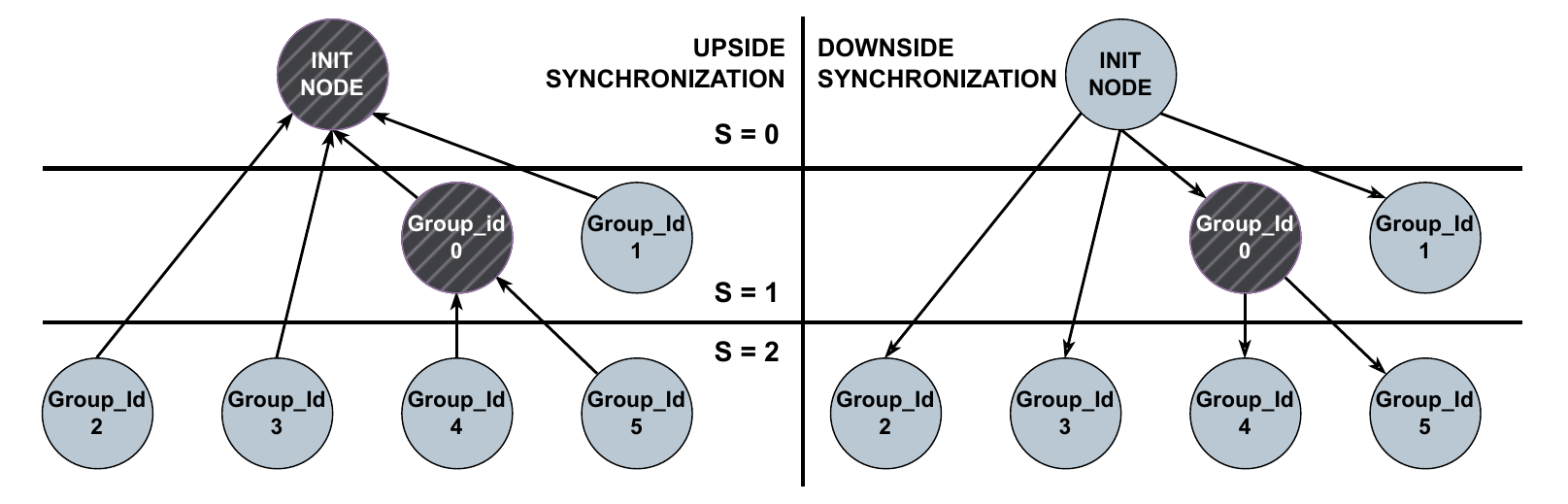}
  \caption[Synchronization of $6$ spawned groups and the initial node]{Synchronization of $6$ spawned groups and the initial node. The arrows indicate the direction of messages, and the stripes the necessity of using Barriers.}
  \label{fig:5-Parallel_Spawn-Synch}
\end{figure*}

The left and right sides of Figure~\ref{fig:5-Parallel_Spawn-Synch} illustrate, respectively, the upside synchronization and downside synchronization steps. In this example, six process groups are spawned in two spawning steps. The arrows indicate which processes send messages to signal their readiness during the \textit{upside} phase, and to confirm that all groups are ready in the \textit{downside} phase. A group represented in black with stripes indicates that it must wait to receive all messages before proceeding. At that point, a \texttt{MPI\_Barrier} is used to synchronize the group, after which the execution can continue.

Listing~\ref{code:5-synch} shows the implementation of the synchronization mechanism, where the first stage is shown in lines L13-L17, the second from lines L19-L28, and the third from lines L30-L41. The processes use the \texttt{world\_c} communicator, which corresponds to \gls{mcw} for spawned processes or to the global communicator among source processes. Then, the \texttt{parent\_c} communicator is an inter-communicator between a spawned group and its parent. Additionally, the variable \texttt{qty\_c} indicates the number of child groups spawned by the calling process, and the corresponding inter-communicators are stored in the \texttt{spawn\_c} vector. A particular \texttt{MPI\_Barrier} function is needed at line L24, because other ranks of the same group could also have children, and therefore the group has to ensure all messages have been received before continuing. This is represented for the \textit{Init Node} group and the \textit{Group\_Id 0} group in the upside synchronization in Figure~\ref{fig:5-Parallel_Spawn-Synch}.

\input{Codes/SynchChildren}

\subsection{Binary connection}
\label{subsec:5-conn_step}
Once synchronization between all groups is completed, the connection operation can begin. The adopted strategy consists of establishing binary connections between pairs of groups in several successive steps. In each one, groups with an identifier less than half the number of active groups perform an \texttt{MPI\_Comm\_accept}, while those with a higher identifier execute an \texttt{MPI\_Comm\_connect}. In the case of an odd number of groups, the middle group is temporarily excluded from that step and proceeds directly to the next one.

Once the connection is completed, both groups are merged into a new communicator and adopt a common group identifier, corresponding to the group that executed the \texttt{MPI\_Comm\_accept} call.

\begin{figure}[tb!]
  \centering
  \includegraphics[width=0.60\linewidth]{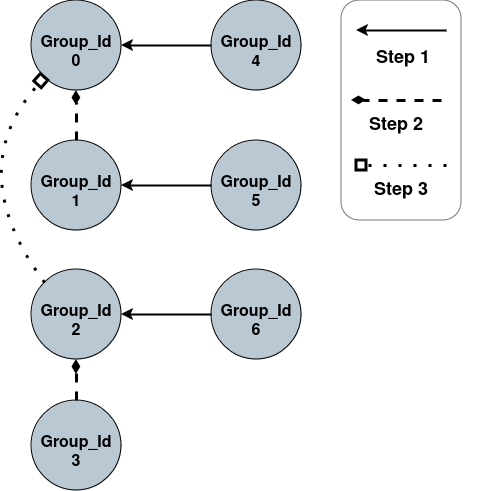}
  \caption[Connection of $7$ spawned groups in $3$ steps]{Connection of $7$ spawned groups in $3$ steps. In each step the number of groups is halved and connections operations may be unordered.}
  \label{fig:5-Parallel_Spawn-Connect}
\end{figure}

This procedure is repeated, halving the number of active groups at each step, until all processes are connected within a single group. Figure~\ref{fig:5-Parallel_Spawn-Connect} shows an example in which seven spawned groups are connected in three steps, continuing the scenario previously illustrated in Figure~\ref{fig:5-Parallel_Spawn-8Nodes}.

Thanks to the synchronization mechanism described earlier, no strict ordering is required for the execution of Connect operations. Groups performing accept simply wait for incoming connections, regardless of the order in which they arrive.

Listing~\ref{code:5-binary_connection} presents a simplified version of the code used to implement this binary connection strategy between groups. In the code, the groups are distinguished based on whether they should perform an \texttt{MPI\_Comm\_accept} operation (lines L15–L19) or an \texttt{MPI\_Comm\_connect} (lines L20–L26). In the latter case, the group obtains the corresponding port name by calling the \texttt{get\_remote\_port} function (line L22), which uses the target group’s \textit{group\_id} to return the port name. To do so, it first writes an established service name based on the \textit{group\_id}, and then calls \texttt{MPI\_Lookup\_name}, which retrieves the port name associated with the service.

\input{Codes/BinaryConnection}

\subsection{Rank Reordering}
\label{subsec:5-reorder_step}

Once all the generated groups have been merged into a single one, the \gls{mpi} ranks may not be ordered across nodes. This is because binary connections between groups do not enforce a specific order and, therefore, are susceptible to race conditions.

To ensure a consistent global order, an \texttt{MPI\_Comm\_split} call is made in which each process specifies its new rank as follows: 
\begin{equation}
    \text{world\_rank} + \sum\limits_{j=0}^{N-1} R_{j} + \sum\limits_{j=0}^{\text{group\_id-1}} S_{j},
    \label{equation:5-reorder}
\end{equation}
where, \textit{world\_rank} is the rank that the calling process had in its \gls{mcw}, while both summations take into account all the previous ranks. The first one counts the number of ranks that existed before the resize, but is not computed since the value is obtained when the group is spawned and can be used directly. Meanwhile, the second summation counts the number of ranks in all groups whose \textit{group\_id} is smaller than that of the calling rank.



\subsection{Overview}
\label{subsec:5-overview}
To conclude this subsection, a global overview of the tasks required to perform the parallel spawn is provided, both for \textit{source} and newly spawned processes. 

The steps for \textit{sources} must be carried out in the following order:
\begin{enumerate}
    \item Open a port and service name for the root rank.
    \item Use one of the strategies described in Subsections~\ref{subsec:5-hypercube} and~\ref{subsec:5-iterative_diff} to perform the parallel spawn.
    \item Synchronize ranks to ensure correctness during the binary connection. (Subsection~\ref{subsec:5-synch_step})
    \item Connect the merged-groups communicator with the initial group communicator.
\end{enumerate}

Listing~\ref{code:5-summary_parents} shows these tasks for the \textit{source} processes. First, they compute the required initial data (line L14), and then the root process opens a port (line L17). At this point, the parallel spawn begins, selecting the appropriate strategy depending on whether the new node allocation is homogeneous or heterogeneous (lines L20–L24). Note that, each algorithm determines whether a given rank must spawn additional groups. Next, the ranks synchronize to ensure correctness (line L27). Afterwards, unnecessary communicators are disconnected (lines L30-L31), and finally the initial group waits for the merged group of spawned processes to connect to the port opened by the root process (lines L32-L34).

In the case of newly spawned processes, the steps to carry out are the following:
\begin{enumerate}
    \item Open a port for root ranks whose group identifier satisfies that ${group\_id} < {total\_spawned\_groups} / 2$. (Prior preparation to Subsection~\ref{subsec:5-conn_step})
    \item Use one of the strategies described in Subsections~\ref{subsec:5-hypercube} and~\ref{subsec:5-iterative_diff} to perform the parallel spawn.
    \item Synchronize ranks to ensure correctness during the binary connection. (Subsection~\ref{subsec:5-synch_step})
    \item Carry out the binary connection to merge the spawned groups into a single one. (Subsection~\ref{subsec:5-conn_step})
    \item Reorder the ranks to match the expected ordering. (Subsection~\ref{subsec:5-reorder_step})
    \item Connect the merged-groups communicator with the initial group communicator.
\end{enumerate}

Listing~\ref{code:5-summary_child} shows the tasks carried out for the generated groups. At startup, they first obtain the data they will use, which they receive from their parent group upon creation (lines L17-L18). Then, in each group, the root rank of those groups whose \textit{group\_id} is lower than half of the generated groups (\textit{groups–I}) proceeds to open a port (lines L21–L22). At this point, the parallel spawn begins, selecting the appropriate strategy depending on whether the new node allocation is homogeneous or heterogeneous (lines L25–L29). Once all groups have been created, they synchronize to ensure that all ports have been opened (line L32), and then disconnect any communicators that are no longer needed (lines L33–L36). After this synchronization, all spawned processes are merged into a single communicator (lines L39-L40) and then reordered according to the expected rank ordering (line L41). Finally, the root process of the merged group obtains the port name of the initial group (lines L44-L45) and connects to it (lines L46-L47).

\input{Codes/summary}

In cases of oversubscription, where the number of active processes exceeds the number of cores allocated to the job, both of the proposed strategies can be applied. In this context, the vector $A$ can be adjusted to reflect the expected level of oversubscription on each node. However, for the first strategy, it is necessary to ensure that all non-zero entries of $A$ are equal across nodes.

An additional important consideration appears for the initial process group that starts the application. When the execution begins using more than one node, the resulting \gls{mcw} cannot returns nodes partially. Within the \gls{mam} library, there are two possible approaches to handle this situation:
\begin{itemize}
    \item Detect this condition during \gls{mam} initialization and recreate all processes using the Baseline method together with the parallel spawning strategy described in this work; or
    \item Detect the issue and postpone any corrective action until it is actually required.
\end{itemize}
In this work, the second approach has been adopted. The reason is that the application may never need to reduce its resource usage. If a shrink is requested, the action to be taken depends on the execution history:
\begin{itemize}
    \item No previous expansions have been performed. The application still relies entirely on the initial \gls{mcw}, which cannot be partially returned. Therefore, in case a future shrink is required, \gls{mam} performs a parallel expansion to enable \gls{ts}.
    \item At least one expansion has been performed. Two cases may arise depending on the requested reduction size:
    \begin{itemize}
        \item The number of nodes to be returned is smaller than the original allocation. Only the nodes allocated in previous expansions are returned, preserving the initial \gls{mcw} intact. The issue is postponed.
        \item The number of nodes to be returned is greater than or equal to the original allocation. The entire initial \gls{mcw} can be safely released. If additional nodes must be returned, they come from the expanded set as well. In this scenario, the challenging \gls{mcw} disappears entirely.
    \end{itemize}
    \item The \gls{rms} requests the release of a subset of cores within a node. The excess ranks can be turned into zombie processes, enabling a partial shrink with \gls{zs}~\cite{spawn_methods}.
\end{itemize}

\subsection{Enabling a TS shrink}
\label{subsec:5-ts_overview}
To ensure a \gls{ts} can be used instead of \gls{zs}, some information must be recorded in the root rank of the global communicator and in the root rank of each \gls{mcw}.

The root rank of the global communicator must maintain a structure that contains for each \gls{mcw}, the nodelist where they are executing. This structure enables the root to decide which approach of Subsection~\ref{subsec:5-overview} should be used when shrinking and decide which groups to terminate. Additionally, if the root rank decides that its own group should be terminated: first, it selects as the new root rank the rank whose identifier is the lowest between the ones that will stay after the reconfiguration; and second it must communicate the structure to that new root rank of the global communicator. 

At the \gls{mcw} level, the root rank maintains a structure that records, for each rank in the communicator, whether it is active or marked as a zombie, as well as the node on which it resides. When a partial shrink affecting only a subset of cores within a node is requested, the corresponding ranks are marked as zombies and enter a sleep state. If, instead, all ranks in the \gls{mcw} are marked as zombies, the group transitions to a \gls{ts} operation: the zombie ranks are awakened and all processes in the communicator terminate.

If the \gls{rms} requests the release of all nodes associated with an \gls{mcw}, the group is terminated directly using \gls{ts}. However, if an \gls{mcw} spans multiple nodes and the \gls{rms} requests the release of only a subset of those nodes, \gls{ts} cannot be applied. In this case, ranks cannot be selectively marked as zombies, and the root of the global communicator must fall back to an alternative shrink strategy based on \gls{zs}.

%% file: Codes/SynchChildren.tex
\begin{lstlisting}[float=htbp!, caption={[Synchronization step before binary connection]Synchronization step before binary connection.}, label=code:5-synch]
void common_synch(MPI_Comm world_c, MPI_Comm parent_c, 
    int qty_c, MPI_Comm *spawn_c) {
  // world_c: Built comm for sources, MCW for targets
  // parent_c: Intercomm to contact parent rank if any
  // qty_c: Total children groups for calling rank
  // spawn_c: Children comms array for calling rank
  int rank, i, root = 0;
  // The color separates ranks into two subgroups
  int color = qty_c ? 1 : MPI_UNDEFINED;
  MPI_Request *reqs = malloc(qty_c * sizeof(MPI_Request));
  MPI_Comm synch_ranks;

// Stage 1: Subcommunicator creation
  // Only root ranks or ranks with children synchronize
  // synch_ranks is only used by ranks with children
  MPI_Comm_rank(world_c, &rank);
  MPI_Comm_split(world_c, color, rank, &synch_ranks);

// Stage 2: Upside Synchronization
  // Wait for all children of this group
  for(i=0; i<qty_c; i++) {
    MPI_Irecv(..., root, spawn_c[i], &reqs[i]);
  }
  if(qty_c){ MPI_Waitall(reqs); MPI_Barrier(synch_ranks); }
  if(parent_c != MPI_COMM_NULL && rank == root) { 
    // Send to parent that this group is ready
    MPI_Send(..., root, parent_c); 
  }  
  
// Stage 3: Downside Synchronization
  if(parent_c != MPI_COMM_NULL && rank == root) { 
    // Wait for the other groups readiness
    MPI_Recv(..., root, parent_c);
  }
  // Notify all ranks in the group to continue
  if(parent_c != MPI_COMM_NULL && qty_c) 
    {MPI_Barrier(synch_ranks);}
  // Notify children that all groups are ready
  for(i=0; i<qty_c; i++)
    {MPI_Isend(..., dest=root, spawn_c[i], &reqs[i]);}
  if(qty_c) {MPI_Waitall(reqs);}
  
  if(synch_ranks != MPI_COMM_NULL) 
    {MPI_Comm_disconnect(&synch_ranks);}
}
\end{lstlisting}

%% file: Codes/BinaryConnection.tex
\begin{lstlisting}[float=htbp!, caption={[Binary connection step for the parallel spawn]Binary connection step for the parallel spawn.}, label=code:5-binary_connection]
void binary_connection(int groups, int group_id, 
                char *my_port, MPI_Comm *new_comm) {
  // groups: Total spawned groups.
  // group_id: Group identifier for calling rank
  // my_port: Local port name of calling rank if any
  // new_comm: Output communicator of combined groups
  int middle, new_groups, new_gid, root = 0;
  char *port;
  MPI_Comm merge_comm, aux_comm, intercomm;
  merge_comm = aux_comm = MPI_COMM_WORLD;
  intercomm = MPI_COMM_NULL;
  while(groups > 1) {
    middle = groups / 2;
    new_groups = groups - middle;
    if(group_id < middle) {
      MPI_Comm_accept(my_port, MPI_INFO_NULL, root,
        merge_comm, &intercomm);
      MPI_Intercomm_merge(intercomm, 0, &aux_comm);
      merge_comm = aux_comm;
    } else if(group_id >= new_groups) {
      new_gid = groups - group_id - 1;
      get_remote_port(new_gid, &port);
      MPI_Comm_connect(port, MPI_INFO_NULL, root,
        merge_comm, &intercomm);
      MPI_Intercomm_merge(intercomm, 1, &aux_comm);
      merge_comm = aux_comm; group_id = new_gid;
    } 
    groups = new_groups;
  } 
  *new_comm =  merge_comm;
}
\end{lstlisting}

%% file: Codes/summary.tex
\begin{lstlisting}[float=htbp!, caption={[Overall tasks for source ranks]Overall tasks for source ranks.}, label=code:5-summary_parents]
void parallel_strat_parents(MPI_Comm group_comm, int rank, 
                    MPI_Comm *out_comm) {
  // group_comm: Built comm for sources
  // rank: Identifier of calling process
  // out_comm: New intercomm for sources and spawned
  int qty_c;
  int groups, root;
  int N, *A, *R, *S;
  char *port_name;
  MPI_Comm *spawn_c = NULL;

  qty_c = 0; root = 0;
  // Calculate following data
  get_initial_data(&N,&A,&R,&S, &groups);

  //Open port if rank 0 (root)
  open_port(&port_name, !rank, groups);

  // Decide spawn depending on core counts
  if(check_homogenous_dist(S)) { 
    hypercube(..., &qty_c, &spawn_c);
  } else { 
    diffusive_iterative(..., &qty_c, &spawn_c); 
  }
  
  // Synchronize all ranks for the binary connection
  common_synch(group_comm, MPI_COMM_NULL, qty_c, spawn_c);

  // Create intercomm between sources and children
  for(int i=0; i<qty_c; i++) 
    {MPI_Comm_disconnect(&spawn_c[i]);}
  if(spawn_c != NULL) free(spawn_c);
  MPI_Comm_accept(port_name, MPI_INFO_NULL, root,
    group_comm, out_comm);
}
\end{lstlisting}

\begin{lstlisting}[float=htbp!, caption={[Overall tasks for target ranks]Overall tasks for target ranks.}, label=code:5-summary_child]
void parallel_strat_children(int rank, MPI_Comm *out_comm, 
                        MPI_Comm *newintracomm) {
  // rank: Identifier of calling process
  // out_comm: New intercomm for sources and spawned
  // newintracomm: Output intracomm for spawned ranks
  int i, exp_id, group_id, opening, qty_c;
  int groups, root;
  int N, I, *A, *R, *S;
  int initial_group_id = -1;
  char *port_name, *remote_port_name;
  MPI_Comm *spawn_c = NULL;
  MPI_Comm group_comm = MPI_COMM_WORLD;
  MPI_Comm_get_parent(out_comm);

  qty_c = 0; root = 0;
  // Data is obtained when rank is spawned
  get_initial_data_children(&N,&A,&R,&S, &groups, &I, 
    &group_id);

  // Open port if required for binary connection
  opening = (rank == root && group_id < (groups-I)/2) ?1:0;
  open_port(&port_name, opening, group_id);

  // Decide spawn depending on core counts
  if(check_homogenous_dist(S)) { 
    hypercube_spawn(..., &qty_c, &spawn_c);
  } else {
    diffusive_iterative_spawn(..., &qty_c, &spawn_c);
  }

  // Synchronize all ranks for the binary connection
  common_synch(group_comm, *out_comm, qty_c, spawn_c);
  for(i=0; i<qty_c; i++) 
    { MPI_Comm_disconnect(&spawn_c[i]); }
  if(spawn_c != NULL) free(spawn_c);
  MPI_Comm_disconnect(out_comm);

  // Connect groups and ensure expected rank order
  binary_connection(groups - I, group_id, port_name, 
    newintracomm);
  rank_reorder(newintracomm, N, R, S, rank);

  // Create intercomm between sources and children
  if(rank == root) 
    get_remote_port(initial_group_id, &remote_port_name);
  MPI_Comm_connect(remote_port_name, MPI_INFO_NULL, root, 
    *newintracomm, out_comm);
}
\end{lstlisting}

%% file: Sources/5-Results.tex
\section{Experimental results}
\label{sec:5-results}
The experiments have been designed to address the following questions:
\begin{itemize}
    \item Do the proposed parallel strategies reduce the resize time when expanding in comparison with the Merge method?
    \item How do these strategies behave in terms of resize time during shrink operations when compared to a \gls{ts} method?
\end{itemize}
The first question evaluates whether the proposed strategies provide an improvement over the Merge method. Even if no improvement is observed, it is still necessary to assess whether the additional overhead is justified, given that the primary goal of the parallel strategies is to enable the use of the extremely fast \gls{ts} method for shrinkage. The second question addresses two objectives. Firstly, it quantifies the difference in resize time between the parallel strategies and the \gls{ts} method, highlighting the effectiveness of the latter. Secondly, it enables direct comparison with results previously reported in the literature~\cite{spawn_methods}.

\subsection{\textit{Hardware} and \textit{Software}}
The experiments were conducted on two clusters. The first one, \gls{mn5}\footnote{\url{https://www.bsc.es/supportkc/docs/MareNostrum5/overview}}, used $32$ nodes from the general queue. Each node has two 56-core Intel Xeon 8480 \gls{cpu}s ($3584$ cores in total). This system was employed to analyze the behavior of both spawning strategies described in Section~\ref{sec:5-method}. 

The second cluster, Nasp\footnote{\url{https://www.hpca.uji.es/computing-facilities/}}, consists of two sets of 8 nodes. The first set has nodes with two 10-core Intel Xeon 4210 \gls{cpu}s ($160$ cores total), connected through a 100 \gls{gbi}/s InfiniBand \gls{edr} and a 10 \gls{gbi}/s Ethernet network. The second set uses nodes with 32-core Intel Xeon 6346 \gls{cpu}s ($256$ cores total) connected via 10 \gls{gbi}/s Ethernet. Both sets communicate through a shared 10 \gls{gbi}/s Ethernet link between their switches. This heterogeneous cluster was used to evaluate the effectiveness of the heterogeneous-aware strategy in Section~\ref{subsec:5-iterative_diff}.

All \gls{mn5} experiments were run with MPICH 4.2.0 and CH4:\gls{ofi} (InfiniBand), while NASP experiments were run with MPICH 3.4.3 and CH3:Nemesis (Ethernet).

Proteo was used to evaluate the different approaches. Before reconfiguration, 5 iterations of Monte Carlo Pi computation including one \texttt{MPI\_Allgather} were performed to ensure \gls{mpi} initialization.

For expansions, \gls{mam} was configured using Baseline (B) and Merge (M) methods, each combined with the two strategies described in Section~\ref{sec:5-method}. All these configurations are compared against Merge without strategies, which was shown to be the best expansion method in a previous study~\cite{spawn_methods}.
For shrinking, \gls{mam} was configured to execute B combined with both strategies described in Section~\ref{sec:5-method}, and these configurations were compared against the Merge method without strategies. When shrinking, the Merge method in \gls{mam} corresponds now to the \gls{ts} method\footnote{Previously, it was the \gls{zs} method, as it was not possible to fully terminate zombies.}; therefore, since no processes are spawned, it cannot use the described parallel strategies. Nevertheless, thanks to the use of parallel strategies in previous resizes, it ensures that \gls{ts} will return the nodes to the \gls{rms} instead of creating zombies. Moreover, since the \gls{zs} method provided the best resizing times in a previous work~\cite{spawn_methods}, its counterpart \gls{ts} is used here both to complement the results of that study and to show the benefits of \gls{ts}. Each configuration was executed $20$ times, and the median result was reported for every triplet of \textit{sources}, \textit{targets}, and method.

Both the Proteo version\footnote{\url{https://lorca.act.uji.es/gitlab/martini/malleability_benchmark/-/tree/Paper-ParallelSpawn}} and the experimental results~\cite{martin_alvarez_2025_17315810} are publicly available.
 
\subsection{Evaluation on Homogeneous Job Allocation}
Both spawning strategies described in Section~\ref{sec:5-method} were evaluated on the homogeneous cluster \gls{mn5}. In total, $5$ configurations were tested for expansion and $3$ for shrinking. Each configuration involved a single reconfiguration from \gls{ns} to \gls{nt} processes, using up to $42$ nodes combinations from the set $\{1, 2, 4, 8, 16, 24, 32\}$.

\begin{figure}[tb!]
    \centering
    \begin{subfigure}[b]{0.95\columnwidth}
        \centering
        \includegraphics[width=\columnwidth]{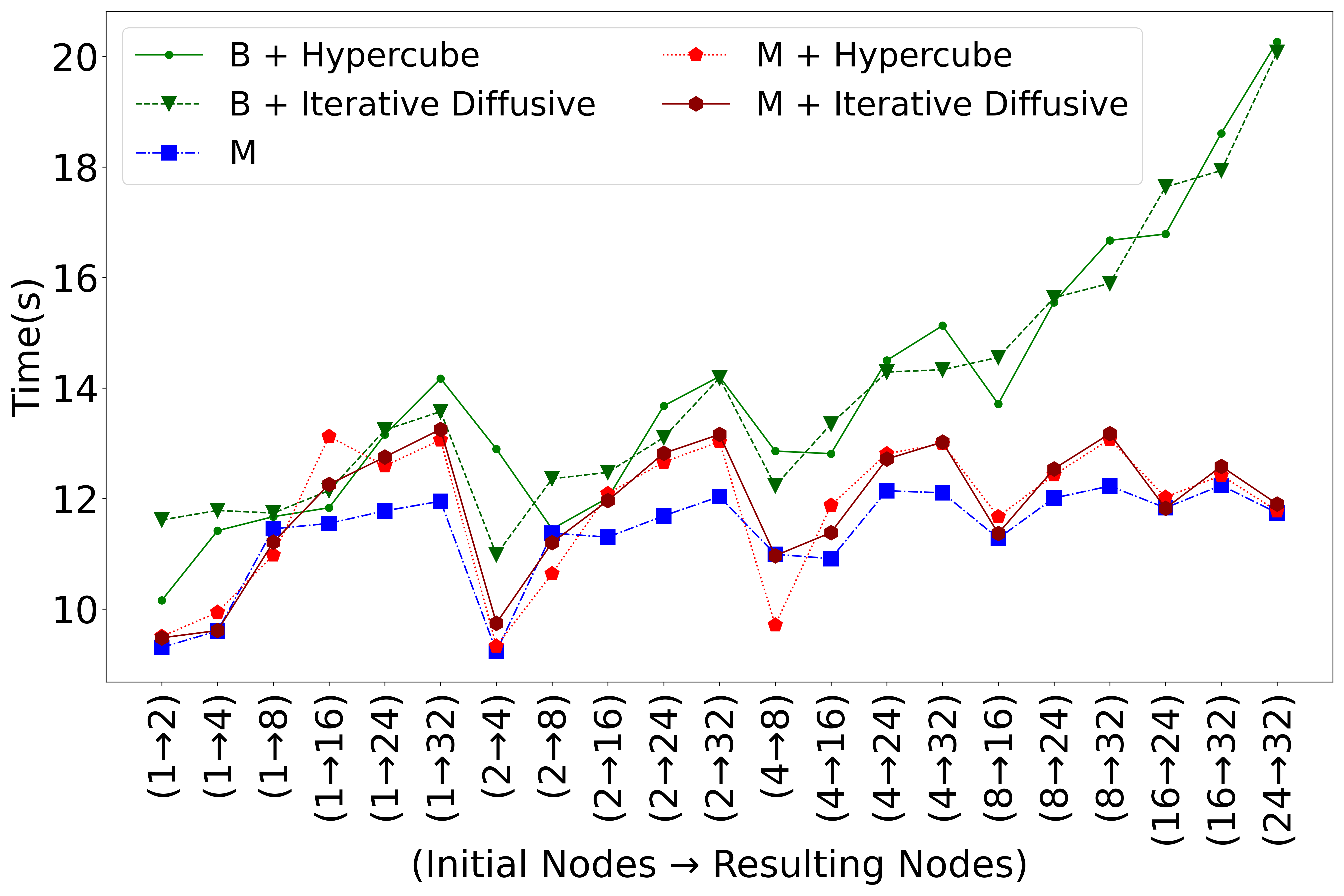}
        \caption{Expansion.}   
        \label{fig:5-mnv_expand}
    \end{subfigure}
    \begin{subfigure}[b]{0.95\columnwidth}
        \centering
        \includegraphics[width=\columnwidth]{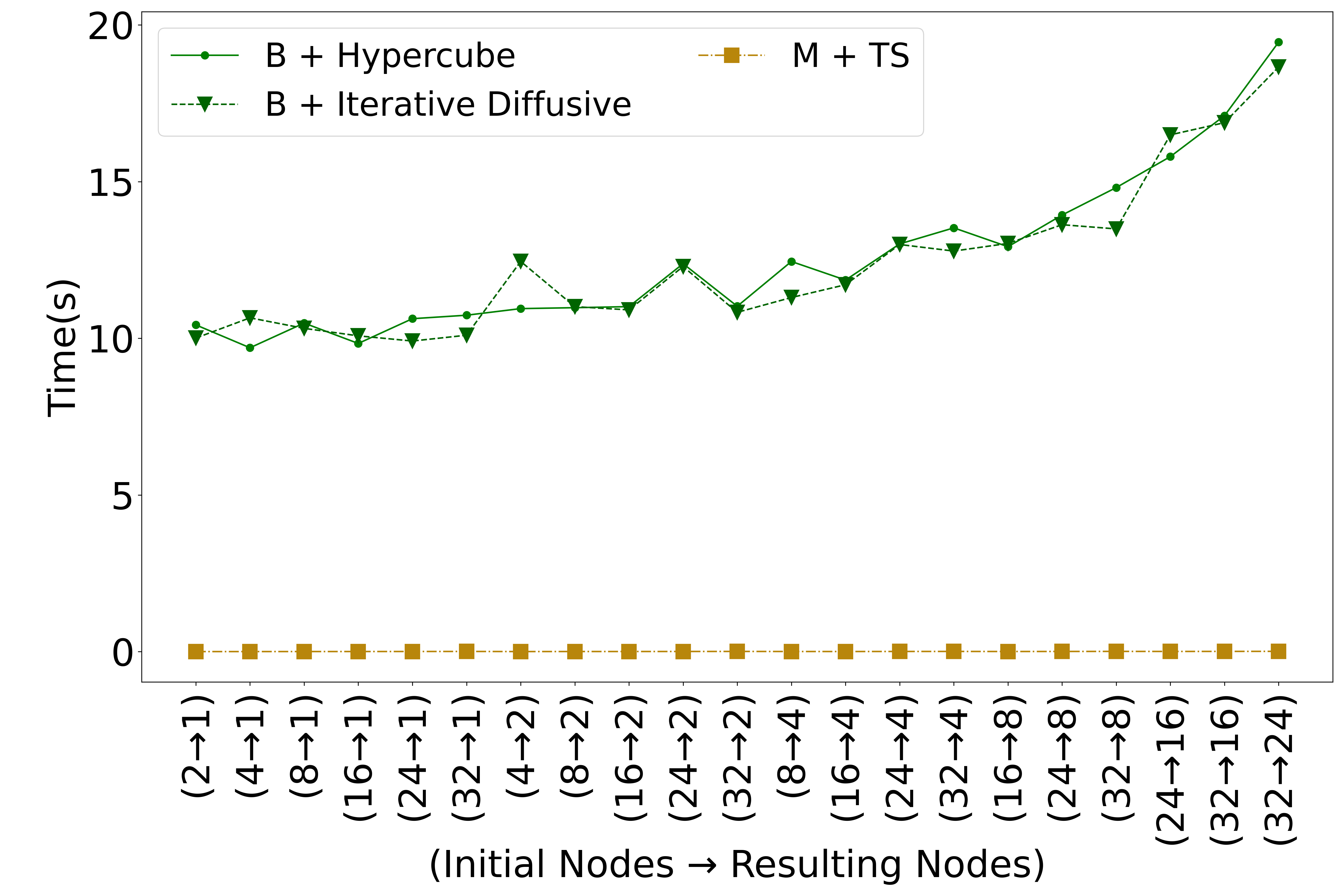}
        \caption{Shrinkage.}   
        \label{fig:5-mnv_shrink}
    \end{subfigure}
    \caption[Resizing times in an homogeneous allocation for MN5]{Resizing times in an homogeneous allocation for \gls{mn5}. The number of processes can be computed by multiplying the nodes by $112$ cores. Letter B refers to the Baseline method, M to Merge method and \gls{ts} to the Merge shrink variant that terminates ranks.}
    \label{fig:5-mnv_results}
\end{figure}

Figure~\ref{fig:5-mnv_results} reports the median reconfiguration times depending on the number of initial nodes ($I$) and resulting nodes ($N$). Figure~\ref{fig:5-mnv_expand} (top) shows expansion results, where the Merge method outperforms all others in $17$ out of $21$ cases ($80.9\%$). Parallel Merge (in both versions) follows closely, requiring up to $1.13\times$ more time due to additional synchronizations among groups. This overhead grows when more than $8$ groups are created and the number of groups is not a power of two. In those cases, at least three steps are required for the binary connection, and the presence of unbalanced leaves increases the cost.

Parallel Baseline methods are consistently slower (up to $1.73\times$), except in one case ($1$ node to $16$ nodes). This overhead stems from spawning extra processes, which also leads to oversubscription.

Figure~\ref{fig:5-mnv_shrink} (bottom) shows shrinking results. In this case, avoiding process spawning is clearly advantageous, resulting in a speedup of $1387\times$ compared to the other evaluated approaches. These findings are consistent with previous work~\cite{spawn_methods}. 

\begin{figure}[tb!]
    \centering
    \includegraphics[width=0.95\columnwidth]{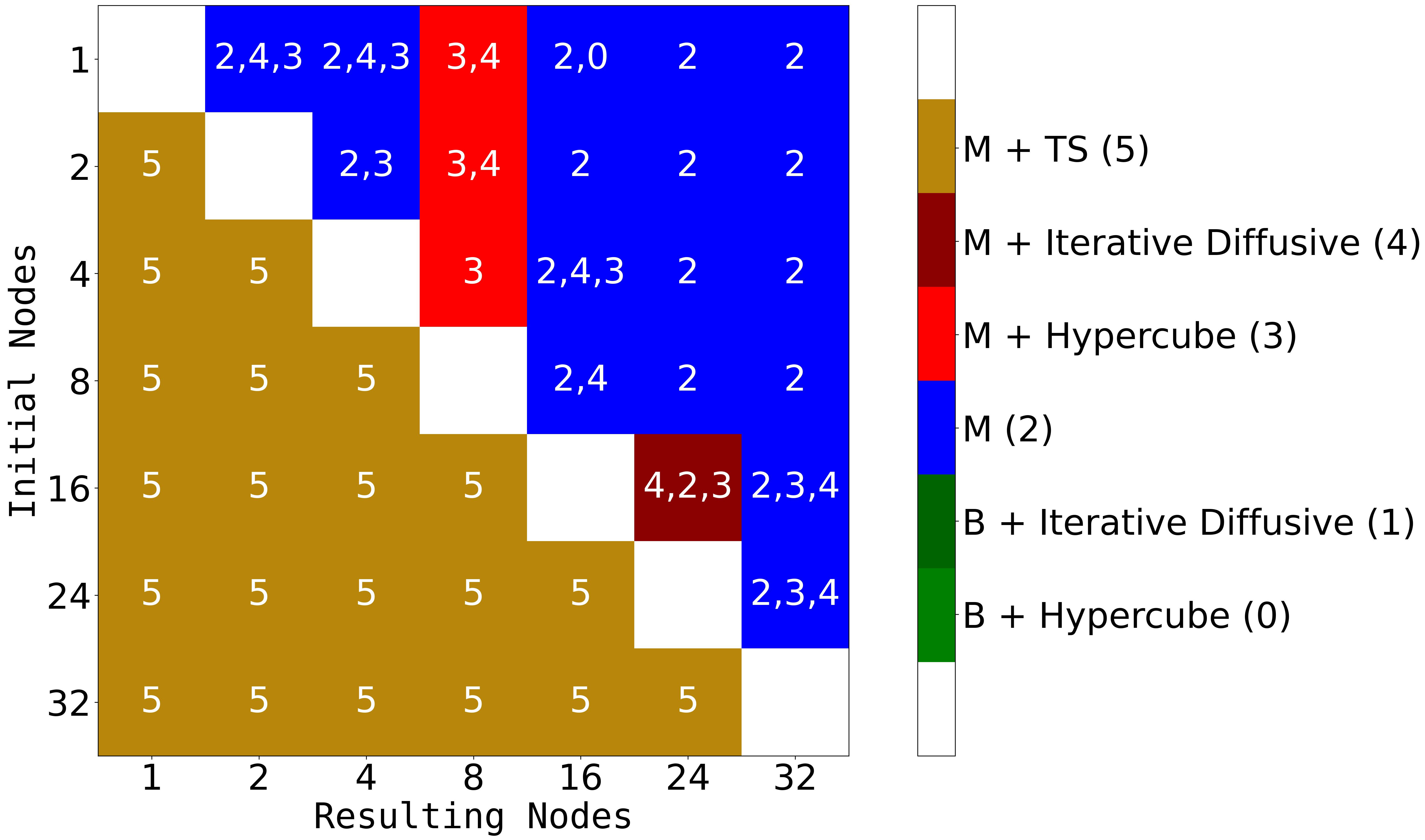}
    \caption[Preferred methods for process management phase depending on the number of nodes $I$ and $N$, in MN5]{Preferred methods for process management phase depending on the number of nodes $I$ and $N$, in \gls{mn5}. The number of processes can be computed by multiplying the nodes by $112$ cores.}
    \label{fig:5-mnv_heatmap}
\end{figure}

Figure~\ref{fig:5-mnv_heatmap} summarizes the best method for each ($I$, $N$) pair based on statistical tests. Axes are defined with \textit{Initial Nodes} on the vertical and \textit{Resulting Nodes} on the horizontal; thus, the upper triangle corresponds to expansion and the lower triangle to shrinkage. When multiple methods appear in a cell, they are statistically equivalent, ordered by ascending time.

The main difference arises in expansions with $8$ or fewer groups, requiring at most $3$ steps for the binary connection. In such cases, the parallel methods are preferred as they enable shrinking without spawning processes, while maintaining times comparable to Merge, though statistically higher.

\subsection{Evaluation of Iterative Diffusive strategy}
The improved strategy described in  Section~\ref{subsec:5-iterative_diff} was evaluated only on the heterogeneous cluster NASP. Three configurations were tested for expansion and two for shrinkage. In this case, the Hypercube strategy is not included in the evaluation because it is unable to correctly spawn the processes. Each configuration involved a single reconfiguration from \gls{ns} to \gls{nt} processes, using up to $72$ combinations of nodes from the set $\{1, 2, 4, 6, 8, 10, 12, 14, 16\}$. In all cases, the number of nodes of each type was balanced (half of one type and half of the other). When only one node was used, the 20-core node was selected.

\begin{figure}[tb!]
    \centering
    \begin{subfigure}[b]{0.95\columnwidth}
        \centering
        \includegraphics[width=\columnwidth]{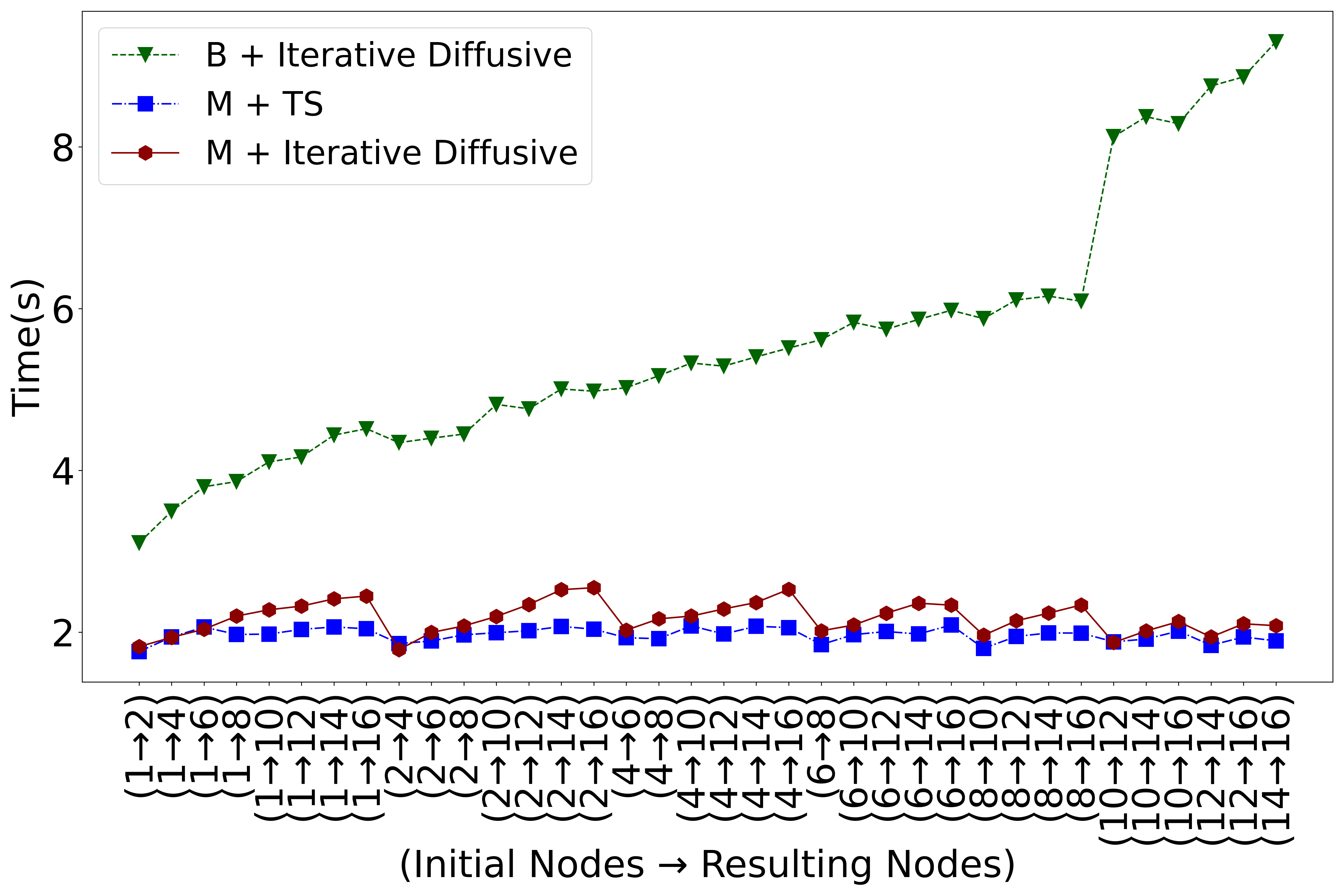}
        \caption{Expansion.}   
        \label{fig:5-nasp_expand}
    \end{subfigure}
    \begin{subfigure}[b]{0.95\columnwidth}
        \centering
        \includegraphics[width=\columnwidth]{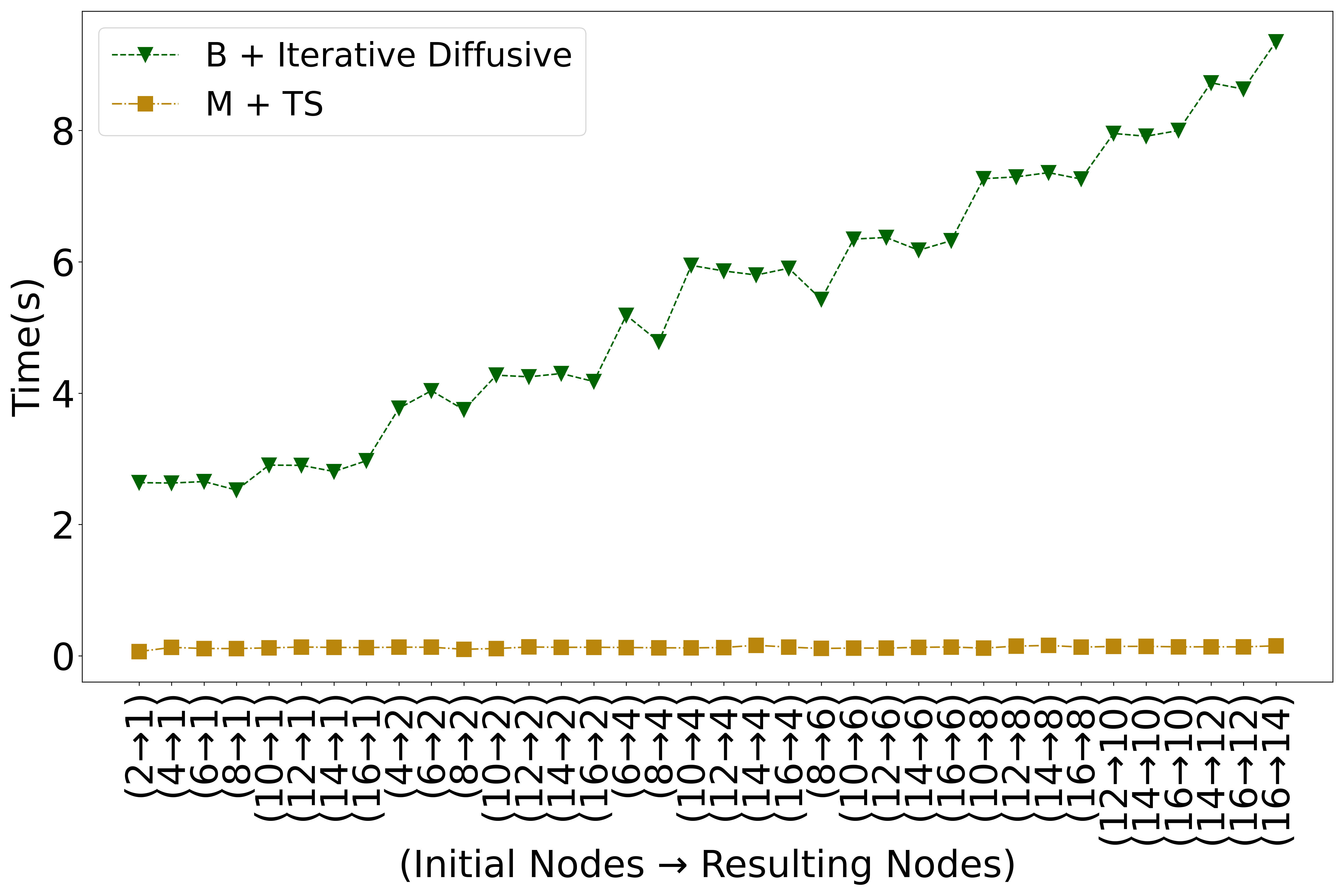}
        \caption{Shrinkage.}   
        \label{fig:5-nasp_shrink}
    \end{subfigure}
    \caption[Resizing times in an heterogeneous allocation for NASP]{
    Resizing times in an heterogeneous allocation for NASP. The number of processes can be computed by multiplying the nodes by $52$ cores, with the exception for the node, where there are $20$ cores. Letter B refers to the Baseline method, M to Merge method and \gls{ts} to the Merge shrink variant that terminates ranks.}
    \label{fig:5-nasp_results}
\end{figure}

Figure~\ref{fig:5-nasp_results} shows the median reconfiguration times as a function of initial nodes ($I$) and resulting nodes ($N$). Figure~\ref{fig:5-nasp_expand} (top) reports expansion results, where the Merge method consistently outperforms the others. The iterative diffusive variant of Merge achieves comparable results, with at most a $1.25\times$ increase, but its cost increases slightly with the number of generated groups.
Baseline methods are always more expensive, ranging from $1.76\times$ to $4.92\times$ slower, due to oversubscription caused by spawning additional processes.

Figure~\ref{fig:5-nasp_shrink} (bottom) presents the shrinking results. As observed in the expansion analysis, the B-based strategies are always the least efficient, as in this case M + \gls{ts} method achieves speedups of at least $20\times$. To reach these benefits a parallel spawn must be performed beforehand, since \gls{ts} requires that each \gls{mcw} to be contained within a single node. Although this parallel spawning for expansions introduces an overhead compared to the previously optimal spawning method~\cite{spawn_methods}, the gains in subsequent shrinking operations outweigh that cost.

Furthermore, in $28$ out of $32$ cases ($87.5\%$) for expanding and shrinking, Merge delivered statistically lower times than the iterative diffusive strategy.

%% file: Sources/6-Conclusions.tex
\section{Conclusions and Future Work}
\label{sec:5-conclusions}
This work introduces and evaluates two new strategies for spawning processes in the traditional \gls{mpi}. Both strategies rely on parallel spawning and have two main goals: (1) reducing the cost of process creation, and (2) enabling efficient shrink operations using the Merge method, the fastest available option. To achieve the second goal, each \gls{mcw} must be fully contained within a single node. Previous Merge implementations could not achieve this because processes that should have been terminated during shrinkage remained as zombies preventing node reuse.

Between the two strategies, the Hypercube approach is limited to homogeneous allocations, whereas the Iterative Diffusive strategy can also be used for resource allocations that use nodes with an heterogeneous number of cores.

Experimental results on both homogeneous and heterogeneous allocations on different systems show that, in the worst case, the expansion with parallel approaches require up to $1.25\times$ times against the Merge method, but they enable shrink operations with a \gls{ts} method that is at least $20\times$ faster than the Baseline method. In summary, this work provides a practical solution for enabling fast shrink operations without compromising expansion performance.

Future work includes studying how to further reduce the synchronization and connection (Subsections~\ref{subsec:5-synch_step} and~\ref{subsec:5-conn_step}) overheads, with the aim of outperforming even the basic Merge expansions. 
Additionally, the load balancing challenges associated with heterogeneous hardware have not been considered in this work and must be addressed.
Another promising direction is the design of data redistribution strategies that minimize transfers, ensuring that processes retain as much data as possible after resizing.

%% file: bib.bib
@misc{osti_1222713,
  author       = {Lucas, Robert and Ang, James and Bergman, Keren and Borkar, Shekhar and Carlson, William and Carrington, Laura and Chiu, George and Colwell, Robert and Dally, William and Dongarra, Jack and others},
  title        = {{DOE Advanced Scientific Computing Advisory Subcommittee (ASCAC) Report:  Top Ten Exascale Research Challenges}},
  url          = {https://www.osti.gov/biblio/1222713},
  place        = {United States},
  year         = {2014},
  month        = {02}
}

@article{project-exascale,
author = {Bernholdt, David E. and Boehm, Swen and Bosilca, George and Gorentla Venkata, Manjunath and Grant, Ryan E. and Naughton, Thomas and Pritchard, Howard P. and Schulz, Martin and Vallee, Geoffroy R.},
title = "{A Survey of MPI Usage in the US exascale computing project}",
journal = {Concurrency and Computation: Practice and Experience},
volume = {32},
number = {3},
pages = {e4851},
url = {https://onlinelibrary.wiley.com/doi/abs/10.1002/cpe.4851},
note = {e4851 cpe.4851},
year = {2020},
}

@InProceedings{Jie2023,
author="Li, Jie
and Michelogiannakis, George
and Cook, Brandon
and Cooray, Dulanya
and Chen, Yong",
title="{Analyzing Resource Utilization in an HPC System: A Case Study of NERSC's Perlmutter}",
booktitle="High Performance Computing",
year="2023",
publisher="Springer Nature Switzerland",
address="Cham",
pages="297--316",
isbn="978-3-031-32041-5"
}

@article{HORI2021102853,
title = {{An international survey on MPI users}},
journal = {Parallel Computing},
volume = {108},
pages = {102853},
year = {2021},
issn = {0167-8191},
url = {https://doi.org/10.1016/j.parco.2021.102853},
author = {Atsushi Hori and Emmanuel Jeannot and George Bosilca and Takahiro Ogura and Balazs Gerofi and Jie Yin and Yutaka Ishikawa},
}

@article{Chadha2021,
  author       = {Mohak Chadha and
                  Jophin John and
                  Michael Gerndt},
  title        = {{Extending {SLURM} for Dynamic Resource-Aware Adaptive Batch Scheduling}},
  journal      = {CoRR},
  volume       = {abs/2009.08289},
  year         = {2020},
  bibsource    = {dblp computer science bibliography, https://dblp.org}
}

@phdthesis{sergiothesis,
address = {Castell{\'{o}} de la Plana},
author = {Iserte, Sergio},
month = nov,
school = {Universitat Jaume I},
title = {{High-throughput Computation through Efficient Resource Management}},
year = {2018},
country = {Spain},
}

@article{Iserte2019a,
author = {Iserte, Sergio and Rojek, Krzysztof},
issn = {0920-8542},
journal = {Journal of Supercomputing},
month = oct,
pages = {1--20},
publisher = {Springer US},
title = {{A Study of the Effect of Process Malleability in the Energy Efficiency on GPU-based Clusters}},
year = {2019}
}

@InProceedings{Cascajo2023,
author="Cascajo, Alberto
and Arbe, Alvaro
and Garcia-Blas, Javier
and Carretero, Jesus
and Singh, David E.",
title="{Malleable Techniques and Resource Scheduling to Improve Energy Efficiency in Parallel Applications}",
booktitle="High Performance Computing",
year="2023",
publisher="Springer Nature Switzerland",
address="Cham",
pages="16--27",
isbn="978-3-031-40843-4"
}

@InProceedings{Sanchez2023,
author="Sanchez-Gallegos, Genaro
and Garcia-Blas, Javier
and Petre, Cosmin
and Carretero, Jesus",
title="{Malleable and Adaptive Ad-Hoc File System for Data Intensive Workloads in HPC Applications}",
booktitle="High Performance Computing",
year="2023",
publisher="Springer Nature Switzerland",
address="Cham",
pages="56--67",
isbn="978-3-031-40843-4"
}

@inproceedings{Dominik22,
    author = {Huber, Dominik and Streubel, Maximilian and Compr\'{e}s, Isa\'{\i}as and Schulz, Martin and Schreiber, Martin and Pritchard, Howard},
    title = {{Towards Dynamic Resource Management with MPI Sessions and PMIx}},
    year = {2022},
    isbn = {9781450397995},
    publisher = {Association for Computing Machinery},
    address = {New York, NY, USA},
    url = {https://doi.org/10.1145/3555819.3555856},
    booktitle = {Proceedings of the 29th European MPI Users' Group Meeting},
    pages = {57–67},
    numpages = {11},
    location = {Chattanooga, TN, USA},
    series = {EuroMPI/USA '22}
}

@article{proteo_2024,
	title = {{Proteo: A Framework for the Generation and Evaluation of Malleable {MPI} Applications}},
	journal = {The Journal of Supercomputing},
	author = {Martín-Álvarez, Iker and Aliaga, Jose I. and Castillo, Maribel and Iserte, Sergio},
	year = {2024},
    pages = {23083-23119},
    issn = {1573-0484},
    url = {https://doi.org/10.1007/s11227-024-06277-5}
}

@inproceedings{mam_api_2024,
author = {Mart\'{\i}n \'{A}lvarez, Iker and Aliaga, Jos\'{e} I. and Castillo, Maribel and Iserte, Sergio},
title = {{MaM: A User-Friendly Interface to Incorporate Malleability into {MPI} Applications}},
year = {2025},
isbn = {978-3-031-90200-0},
publisher = {Springer Nature Switzerland},
address = {Cham},
url = {https://doi.org/10.1007/978-3-031-90200-0_28},
booktitle = {Euro-Par 2024: Parallel Processing Workshops},
pages = {346--358},
numpages = {12},
location = {Madrid, Spain},
series = {}
}

@article{spawn_methods,
    author = {Iker Martín-Álvarez and José I Aliaga and Maribel Castillo and Sergio Iserte and Rafael Mayo},
    title ={{Dynamic Spawning of MPI Processes Applied to Malleability}},
    journal = {The International Journal of High Performance Computing Applications},
    volume = {38},
    number = {2},
    pages = {69-93},
    year = {2024},
    URL = { https://doi.org/10.1177/10943420231176527},
}

@Article{State22,
AUTHOR = {Aliaga, Jose I. and Castillo, Maribel and Iserte, Sergio and Martín-Álvarez, Iker and Mayo, Rafael},
TITLE = "{A Survey on Malleability Solutions for High-Performance Distributed Computing}",
JOURNAL = {Applied Sciences},
VOLUME = {12},
YEAR = {2022},
NUMBER = {10},
ARTICLE-NUMBER = {5231},
URL = {https://www.mdpi.com/2076-3417/12/10/5231},
ISSN = {2076-3417},
}

@misc{martin_alvarez_2025_17315810,
  author       = {Martín Álvarez, Iker},
  title        = {{Proteo Dataset (2025) for an study for Parallel
                   Spawning in dynamic resources
                  }},
  month        = oct,
  year         = 2025,
  publisher    = {Zenodo},
  version      = {1.0.0},
  url          = {https://doi.org/10.5281/zenodo.17315810},
}

@article{State24,
	title = {{Malleability in Modern {HPC} Systems: Current Experiences, Challenges, and Future Opportunities}},
	issn = {1558-2183},
	pages = {1--14},
	journal = {{IEEE} Transactions on Parallel and Distributed Systems},
	author = {Tarraf, Ahmad and Schreiber, Martin and Cascajo, Alberto and Besnard, Jean-Baptiste and Vef, Marc-André and Huber, Dominik and Happ, Sonja and Brinkmann, André and Singh, David E. and Hoppe, Hans-Christian and Miranda, Alberto and Peña, Antonio J. and Machado, Rui and Gasulla, Marta Garcia- and Schulz, Martin and Carpenter, Paul and Pickartz, Simon and Rotaru, Tiberiu and Iserte, Sergio and Lopez, Victor and Ejarque, Jorge and Sirwani, Heena and Wolf, Felix},
	date = {2024-06},
    year = {2024},
}

@inproceedings{Martin2013,
author = {Mart{\'{i}}n, Gonzalo and Marinescu, Maria-Cristina and Singh, David E. and Carretero, Jes{\'{u}}s},
booktitle = {Euro-Par Parallel Processing},
isbn = {978-3-642-40046-9},
month = "aug",
pages = {138--149},
title = {{FLEX-MPI: an MPI Extension for Supporting Dynamic Load Balancing on Heterogeneous Non-dedicated Systems}},
year = {2013}
}

@misc{mpich_website, 
 title={{MPICH Website}}, 
 url={https://www.mpich.org/},
 journal={Mpich}, 
 author="{MPICH Development team}",
 year="2001"
}

@INPROCEEDINGS{Flex_mpi_spawning,
  author={Muñoz, Javier Fernández and Cascajo García, Alberto and Pérez, Jesús Carretero},
  booktitle={2023 IEEE 29th International Conference on Parallel and Distributed Systems (ICPADS)}, 
  title={{Dynamic management of processes and communicators in malleable MPI applications}}, 
  year={2023},
  volume={},
  number={},
  pages={848-855},
  url={https://doi.org/10.1109/ICPADS60453.2023.00127}
}

@incollection{Feitelson1996,
author = {Feitelson, Dror G.},
booktitle = {Lecture Notes in Computer Science book series (LNCS, volume 1162)},
pages = {89--110},
publisher = {Springer, Berlin, Heidelberg},
address = {Heidelberg, Germany},
title = {{Packing schemes for gang scheduling}},
year = {1996}
}

@article{ISERTE2025a,
title = {{Resource optimization with MPI process malleability for dynamic workloads in HPC clusters}},
journal = {Future Generation Computer Systems},
volume = {174},
pages = {107949},
year = {2026},
issn = {0167-739X},
url = {https://www.sciencedirect.com/science/article/pii/S0167739X25002444},
author = {Sergio Iserte and Iker Martín-Álvarez and Krzysztof Rojek and José I. Aliaga and Maribel Castillo and Weronika Folwarska and Antonio J. Peña},
}

@INPROCEEDINGS{Dominik25,
  author={Huber, Dominik and Iserte, Sergio and Schreiber, Martin and Peña, Antonio J. and Schulz, Martin},
  booktitle={ISC High Performance 2025 Research Paper Proceedings (40th International Conference)}, 
  title={{Bridging the Gap Between Genericity and Programmability of Dynamic Resources in HPC}}, 
  year={2025},
  volume={},
  number={},
  pages={1-11},
}
